\documentclass[10pt,prb,nobibnotes,showpacs,showkeys,superscriptaddress,maxnames=6]{revtex4-1}
\usepackage{geometry}
\usepackage{color}
\usepackage{amsfonts,amsmath,amssymb,graphicx, placeins, bm}
\usepackage{setspace,color}

\usepackage{ae,aeguill}
\usepackage[utf8]{inputenc}    
\usepackage[T1]{fontenc}
\newcommand{\ddi}{\hat{\mu}_i\hat{\mu}_j -3 (\hat{\mu}_i \hat{r}_{ij})(\hat{\mu}_j \hat{r}_{ij})}

\newcommand{\stl}{\vskip 0.04\textheight}
  \newcommand{\be}{\beta}     
\newcommand{\la}{\lambda} \newcommand{\al}{\alpha}   
\newcommand{\gD}{\Delta}  \newcommand{\ep}{\epsilon} \newcommand{\s}{\sigma}
     
 \newcommand{\tend}{\rightarrow}
\newcommand{\equa}[1]{\begin{eqnarray} \label{#1}}
\newcommand{\auqe}{\end{eqnarray}}
\newcommand{\equab}[1]{\begin{widetext}\begin{eqnarray} \label{#1}}
\newcommand{\auqeb}{\end{eqnarray}\end{widetext}}
\newcommand{\tab}[1]{\begin{tabular}{#1}}
\newcommand{\bat}{\end{tabular} \\ }
\begin {document}
  \title {Ferromagnetic frozen structures from the dipolar hard spheres fluid
 at moderate and small volume fractions.}
  \author {J.-G. Malherbe}
  \email [e-mail address: ] {malherbe@u-pec.fr}
  \affiliation{ICMPE, UMR 7182 CNRS and UPEC 2-8 rue Henri Dunant 94320 Thiais, France.}
  \author {V. Russier}
  \email [e-mail address: ] {vincent.russier@cnrs.fr}
  \affiliation{ICMPE, UMR 7182 CNRS and UPEC 2-8 rue Henri Dunant 94320 Thiais, France.}
  \author{Juan J. Alonso}
  \email [e-mail address: ] {jjalonso@uma.es}
  \affiliation{F\'{\i}sica Aplicada I, Universidad de M\'alaga, 29071 M\'alaga, Spain}
  \affiliation{Instituto Carlos I de F\'{\i}sica Te\'orica y Computacional,  Universidad de M\'alaga, 29071 M\'alaga, Spain}

  \begin {abstract}
~ \stl
We study the magnetic phase diagram of an ensemble of dipolar hard spheres (DHS) with or without 
uniaxial anisotropy and frozen in position on a disordered structure by tempered Monte Carlo simulations.
The crucial point is to consider an anisotropic structure, obtained from the liquid state of 
the dipolar hard spheres fluid, frozen in its polarized state at low temperature.
The freezing inverse temperature $\be_f$ determines the degree of anisotropy
of the structure which is quantified through a structural nematic order parameter, $\la_s$.
The case of the non zero uniaxial anisotropy is considered only in its infinitely
strong strength limit where the system transforms in a dipolar Ising model (DIM).
The important finding of this work is that both the DHS and the DIM with a frozen structure build 
in this way present a ferromagnetic phase at volume fractions below the
threshold value where the corresponding isotropic DHS systems  exhibit a spin glass
phase at low temperature.
  \end {abstract}
  \maketitle
\section {Introduction}
\label {intro}

Magnetic nanoparticle (MNP) assembled in densely packed structures whether organized or not in large 
scale supracrystals still arouse a great interest both for their potential applications and the 
fundamental point of view. 
In these superstructures the interactions between neighboring MNP at high concentration lead to 
strong coupling configurations and thus to collective behaviors resulting in new (specific) or
enhanced properties. 
The understanding of the onset of these collective behaviors, such as the complex magnetic states,
and the interplay with the underlying structure of the MNP still necessitates modeling work
to clarify some points as for instance the nature of the magnetic phases in MNP superstructures.
One is faced to quite different situations according to the physical state of the MNP assemblies
in ferrofluids or supracrystals where the MNP are either freely moving in liquid state or frozen 
in solid structures respectively. The link between these two states is important as experimentally the 
latter is obtained from the former by the solvent evaporation~\cite{josten_2017,costanzo_2020}.
For small enough MNP falling in the single domain regime, the modeling of the MNP assembly can be performed 
in the framework of the effective one spin (or macro spin) model, thus avoiding the multi scale character 
resulting from the internal structure of the MNP. \\ \indent
At the macro-spin level, the interaction effects in MNP assemblies can be described
by an ensemble of particles bearing a constant magnetic moment interacting through the dipole dipole
interaction (DDI) and a short range potential and
undergoing both the magnetocrystalline anisotropy (MAE) and the external field. 
When the short range contribution to the interaction is reduced to the hard sphere
potential describing the steric effects, one gets the widely studied dipolar hard sphere fluid (DHS)
at least in the liquid state if one considers that since the MNP are free to rotate, the easy axes follow 
instantaneously the moments and the MAE can be ignored.
Both the dipolar hard sphere fluid at sufficiently high density~\cite{wei_1992a} and the ensemble of dipoles
frozen on a perfect 
face centered cubic (FCC) or body centered tetragonal (BCT) lattice~\cite{bouchaud_1993,weis_1993} 
in the absence of MAE present a 
well defined ferromagnetic phase at low temperature. This is also the case for an ensemble of dipolar 
hard spheres frozen in a random hard sphere like distribution with volume fraction larger than 
$\Phi_s\simeq{0.495}$~\cite{alonso_2020b}.
In presence of MAE or with higher dilution the magnetic phase diagram of the 
frozen MNP superstructures is mainly determined by the competition
between the DDI induced 
ferromagnetic (FM) order and the disorder stemming from the structure and/or the MAE
through either its magnitude or the easy axes distribution. 
\\ \indent
This order/disorder competition as the driving force of the phase diagram is a general feature
and is found also for Ising or Heisenberg models with short range interactions. 
The Ising model with randomly distributed $\pm{J}$ bonds on simple cubic (SC) lattice has been investigated 
in details~\cite{hasenbusch_2007,papakonstantinou_2013,papakonstantinou_2015}.
In this model, the disorder control parameter is the probability $p$ of anti-ferromagnetic coupling, $J_{ij}=-J$.
%
The phase diagram of the Heisenberg model, also with interactions limited to nearest neighbors and including the 
uniaxial MAE with random distribution of easy axes, the so-called random anisotropy model introduced by 
Harris~{\it et al.}~\cite{harris_1973} has been investigated in Ref~\cite{itakura_2003}. 
Here the disorder strength is the anisotropy over exchange coupling ratio, $D/J$. 
As in the case for the Ising $\pm{J}$ models by increasing the value of $D/J$
the ordered phase at low temperature transforms gradually from a FM to a 
spin-glass (SG) state~\cite{nguyen_2009a}. 
Moreover the special case $D/J=4$ has been investigated 
in Ref.~\cite{nguyen_2009b} with an isotropic random or a cubic anisotropic random distribution of easy axes 
confirming in these two cases the FM quasi long range order (QLRO-FM) state at low temperature. 
\\ \indent
The magnetic phase diagram of the DHS for a frozen disordered isotropic structure was first investigated 
in a mean field approximation~\cite{zhang_1995} and then by Ayton {\it et al}~\cite{ayton_1995,ayton_1997} where, 
using a high temperature liquid free of DDI to build the distribution of particles, the frozen DHS is found 
to order in a dipolar glass instead of a FM phase at low temperature and for a volume fraction of $c.a.$~0.42. 
This phase diagram for different situations of DHS frozen 
distributions~\cite{alonso_2019,russier_2020a,russier_2020b,alonso_2020b} with or without MAE has been revisited 
in more details recently. In each of the situations considered the structure is disordered and isotropic 
or ordered on a lattice with cubic symmetry and there is one disorder control parameter, say $x$. 
The phase diagram is then determined in the $(T,x)$ plane. On the perfect FCC lattice with finite MAE the distribution of
easy axes is random and $x$ is the MAE over DDI coupling ratio, $\la_u$. In the infinite MAE limit where the model reduces
to a dipolar Ising model (DIM) the distribution of easy axes is textured and $x$ is the variance $\s$ of the distribution.
In the case of the frozen isotropic hard sphere (HS) like distribution, 
the infinite MAE limit with textured distribution of easy axes was considered 
with the volume fraction $\Phi$ fixed to its maximum limit, 
the so-called random close packing limit (RCP), $\Phi_{RCP}=0.64$. 
Finally, the system free of MAE was also considered
for frozen isotropic HS distributions.
In this case, where $x$ is 
directly related to the volume fraction $\Phi$, ($x=1-\Phi/\Phi_{RCP}$),
the limit between the FM and SG states was found at $\Phi_s\simeq{}0.495$.
The figure~(\ref{schema}) is an illustrative example of this scheme for the case of DHS for frozen isotropic distribution 
either free of MAE displaying the FM/SG line at $\Phi=\Phi_s$ or in the DIM limit where $x$ is the variance of the axes 
distribution. 
The general rule is that, with the increase of $x$, the ordered phase at low temperature goes from the 
FM to the SG one, with possibly an intermediate quasi long range ordered FM state with transverse SG. 
\\ \indent
Furthermore the FM ordering in the liquid DHS simultaneously breaks the 
symmetry and induces a structure anisotropy ~\cite{wei_1992b,gingras_1995} which leads to the tetragonal-I 
bond orientational order~\cite{wei_1992b}, 
below a solid liquid transition temperature, $T_{sl}$. 
The structure in the liquid DHS FM phase is then characterized by
$g_{\parallel}\neq{}g_{\perp}$, where $g_{\parallel}(r)$ and $g_{\perp}(r)$ are the 
longitudinal and transverse pair distribution functions relative 
to the polarization direction respectively.
\\ \indent
In the present work, 
we propose to make use of this structural
anisotropy to obtain particular frozen configurations of DHS and hence
restore the FM order at low density with respect to the SG transition that 
was observed until now. We thus investigate the case of a frozen DHS whose structure is anisotropic as
obtained from the DHS in the liquid state frozen at a temperature, say $T_{f}$, below its critical temperature $T_c(DHS)$. 
The degree of anisotropy of such a structure increases with the decrease of $T_{f}$ and can be quantified through a 
nematic order parameter, $\la_s$ built from the distribution of first neighbors bonds as already suggested 
in~\cite{gingras_1995}. We consider two values of the volume fraction, one $\Phi=0.45$ is just below the threshold
of the onset of the FM phase in the frozen hard-sphere like isotropic distribution, and the second one, 
$\Phi=0.262$ corresponds to a low density DHS fluid. 
We investigate first the system free of MAE, and in a second step the infinite MAE limit where the model becomes 
a dipolar Ising model (DIM). The main purpose of the present work is to investigate a possible way for the frozen DHS 
to order in a FM state through the anisotropy of the structure, the latter being induced by the DDI in the liquid state.
Hence our second purpose is to make the link between the properties of the ferrofluid with those of the corresponding frozen
superstructures.
\\ \indent
\\ \indent
The paper is organized as follows.
In section~\ref{model} we present the model and the simulation details. In section~\ref{results} the results are presented
starting from the necessary elements of the liquid DHS generating the structures and the frozen DHS and frozen DIM are
then discussed. The section~\ref{conclusion} concludes the paper.
%
 \section {Model}
 \label   {model}
 We place ourselves in the framework of the effective one spin model where each single domain MNP is assumed to be
uniformly magnetized with a temperature independent saturation magnetization $M_s$.
 Hence, to model the assembly of MNP free of super exchange interactions, characterized by a uniaxial magnetocrystalline 
anisotropy (MAE) we 
consider a system of dipolar hard spheres of moment $\vec{\mu}_i=\mu_i\hat{\mu}_i$, with $\mu_i=M_{s}v_i$, 
interacting through the usual dipole dipole interaction (DDI) and subjected to a 
one-body anisotropy energy, $K_{i}v_i(\hat{n}_i\ldotp\hat{\mu}_i)^2$.
$K_i$, $\hat{n}_i$ and $v_i$ are the anisotropy constant, the easy axis and the volume of the particle $i$ respectively.
The MNP ensemble is monosdisperse with MNP diameter $d$. The hamiltonian of the system is given by
\equa{beh_1}
  \be H & = & 
        \frac{1}{2} \be \ep_d \sum_{i\neq j} \frac{\ddi}{(r_{ij}/d)^3}
   - \be K v(d) \sum_i (\hat{n}_i\ldotp\hat{\mu}_i)^2 + \frac{1}{2} \be \sum_{i\neq j} v_{sr}(r_{ij})
    ~~~ \textrm{with}~ \ep_d = \frac{\mu_0}{4\pi}\frac{\mu^2}{d^3} 
\auqe
where $\hat{r}_{ij}$ is the unit vector carried by the vector joining sites $i$ and $j$,
$\be=1/(k_BT)$ is the inverse temperature.
$v_{sr}(r)$ is the short range contribution, taken as the hard sphere potential, 
$v_{sr}(r>d)=0$ and $v_{sr}(r<d)=\infty$.
Here the reduced temperature is chosen as $T^*=Tk_B/\ep_{d}$.
Equation~(\ref{beh_1}) is then rewritten as
\begin{subequations}\label{beh_2}
\equa{beh_2a}
  \be H  &=&  \frac{1}{T^*} \left( \frac{1}{2} \sum_{i\neq j} \frac{\ddi}{(r_{ij}/d)^3} 
	      - \la_u \sum_i (\hat{n}_i\ldotp\hat{\mu}_i)^2  
	      + \frac{1}{2\ep_d} \sum_{i\neq j} v_{sr}(r_{ij}) \right )
   ~~~~ \la_u = \frac{Kv(d)}{\ep_d} 
\auqe
\equa{beh_2b}
    \equiv \frac{1}{T^*} \left( \frac{1}{2} \sum_{i\neq j} 
                        \hat{\mu}_i \bar{T}_{ij}  \hat{\mu}_j 
                        - \la_u \sum_i (\hat{n}_i\ldotp\hat{\mu}_i)^2 
			+ \frac{1}{2\ep_d} \sum_{i\neq j} v_{sr}(r_{ij}) \right )
\auqe
\end {subequations}
which introduces the MAE coupling constant $\la_u$ and the dipolar coupling tensor $\bar{T}_{ij}$.
In the following we will consider two different situations for the MAE term : either $\la_u=0$, 
or $\la_u=\infty$ where the model transforms in a dipolar Ising model (DIM) with Ising axes 
coinciding with the easy axes $\hat{n}_i$ and $\hat{\mu}_i=s_i\hat{n}_i$ with the Ising variables 
$s_i=\pm{1}$. Therefore, in the following the MAE is formally dropped and the dipolar Ising model 
is characterized by the set of coupling constants $J_{ij}=\hat{n}_i\ldotp\bar{T}_{ij}\ldotp\hat{n}_j$

The simulation box is a cube with edge length $L$, the total number of dipoles is $N$ and the volume fraction
is $\Phi=N\pi{}d^3/(6L^3)$.
We consider periodic boundary conditions by repeating the simulation cubic box identically in the 3 dimensions.
The long range DDI interaction is treated through the Ewald summation technique~\cite{allen_1987,weis_1993}, 
with a cut-off $k_c=8\;k_m$, $k_m=(2\pi/L)$, in the sum of reciprocal space and the $\al$ parameter of the 
direct sum chosen is $\al=5.80$~\cite{weis_1993}.
The Ewald sums are performed with the so-called 
conductive external conditions~\cite{allen_1987,weis_1993}, i.e. the system is embedded in a medium with 
infinite permeability, $\mu_s=\infty$, which is a way to avoid the demagnetizing effect and thus 
to simulate the intrinsic bulk material properties regardless of the external surface and system 
shape effects. 

In the following the DHS will refer to the usual liquid state where both the $\{\vec{r}_i\}$ and $\{\hat{\mu}_i\}$ 
are moved in the simulation, while the frozen DHS denotes the model with a frozen distribution of the particles
and free of MAE, $\la_u=0$, and the frozen DIM denotes the model in the $\la_u\tend\infty$ limit with both the
particles and their easy axes frozen, which fixes the set of Ising coupling constants $J_{ij}$.
%
\subsection {Simulation method}
\label{sim_meth}

In order to thermalize in an efficient way our system presenting strongly frustrated states,
we use parallel tempering algorithm~\cite{earl_2005,marinari_1992,hukushima_1996} 
(also called tempered Monte Carlo) for our Monte Carlo simulations. Such a scheme is widely used in similar systems,
and we refer the reader to Refs.~\cite{alonso_2010,alonso_2019,russier_2020a} for the details of
the implementation. 
The method is based on the simultaneous simulation runs of 
identical replica for a set of temperatures $\{T^*_n\}$ with exchange trials of the configurations 
pertaining to different temperatures each $N_M$ Metropolis steps according
to an exchange rule satisfying the detailed balance condition. The set of temperatures is chosen
in such a way that on the first hand it brackets the paramagnetic ordered state transition temperature
and on an other hand it leads to a satisfying rate of exchange between adjacent temperature configurations.
Our set $\{T^*_n\}$ is either an arithmetic or geometric distribution, which appears to be a
good approximation of the one built from the
efficient constant entropy increase method~\cite{sabo_2008}.
The arithmetic distribution range used in the frozen DIM is $T^*\in[0.5,3.5]$ with a spacing
$\gD{}T^*=0.05$ and some additional points in temperature have been introduced, in between 
$T^*=2.0$ and 3.0 with a spacing $\gD{}T^*=0.025$.
The geometrical distribution is used principally for the DHS and frozen DHS at $\Phi=0.45$,
in the temperature range $T^*\in[0.08,0.8]$ with 48 temperatures for the system sizes 
$N=1177$ and $758$ while the smallest system size, $N=453$ we use 32 temperatures. 
The simulations on the BCT lattice with $c=1$ at $\Phi=0.45$ have been performed 
with $T^*\in[0.12,1.3]$ with 64, 48 and 32 temperatures for $N=1536$, 648 and
350 respectively.
At lower packing fraction $\Phi=0.262$, for the DHS the temperature range
$T^*\in[0.1,0.2]$ was explored with  $24$ temperatures for system size $N=182$
and $364$ and with $28$ temperatures with $N=728.$
For the frozen DHS we use 32 temperatures in the range [0.2,\;0.7].
When necessary, precise interpolation for temperatures between the points actually 
simulated are done through reweighting methods~\cite{ferrenberg_1988}.
In the frozen disorder situations, the averaging is performed in two steps from a number $N_r$
of realizations of the corresponding disordered structure, including the easy axes distribution in the
frozen DIM and the mean value of any observable, say $A$, is given by
$[<A>]=(1/N_r)\sum_{r}<A_r>_{T}$, $<.>_T$ being the thermal mean value obtained from the Monte Carlo simulation.
The number of realizations is between 100 and 250 for the frozen DHS and frozen DIM, while it is up to 10 
for the DHS where particles are free to move.
The number of Monte Carlo steps is $5\;10^5$ for $\Phi=0.45$ and more than $10^6$ at low density 
for thermalization and $5\;10^5$ accumulation.
\subsection {Observables}
\label {obs}

Our main purpose is the determination of the transition temperature between the paramagnetic and the ordered 
phase and on the nature of the latter, namely ferromagnetic or spin-glass, in terms of the disorder strength.
For the PM/FM transition, 
we consider the spontaneous magnetization 
\equa{m_1}
   m = \frac{1}{N} \left\Arrowvert  \sum_i \hat{\mu}_i \right\Arrowvert
\auqe
computing its moments, $m_k=[<m^k>]$, k = 1,2 and 4.
We compute also the nematic order parameter $P_2$ together with the instantaneous nematic direction, $\hat{d}$ which are 
the largest eigenvalue and the corresponding eigenvector respectively of the tensor
$  
 \bar{Q} = \frac{1}{N} \sum_i (3\hat{\mu}_i\hat{\mu}_i - \bar{I})/2.
$ 
The spontaneous magnetization can also be studied in the ordered phase from the mean value
projected total magnetization on the nematic direction~\cite{weis_1993}, which defines 
\equa{m_la}
   m_{d} =  \frac{1}{N} \sum_i \hat{\mu}_i.\hat{d} 
\auqe 
We compute the mean value $m_{1d}=[<|m_{d}|>]$ and the moments $m_{nd}=[<m_{d}^n>]$, with $n=2,4$.
To locate the transition temperature, $T^*_c$, as usually done, we will use the finite size scaling (FSS) analysis of the Binder
cumulant which is defined either from the moments $m_k$ or  $m_{kd}$ characterized by 3 or one degree of freedom respectively
 \equa{bm_la}
   B_m = \frac{1}{2} \left( 5 - 3 \frac{m_{4}}{m_2^2} \right) , ~~~~~~~~  
   B_{md} = \frac{1}{2} \left( 3 - \frac{m_{4d}}{m_{2d}^2} \right)
 \auqe
From these normalizations, $B_m,B_{md}\tend{}1$ in the long range FM phase and $B_m,B_{md}\tend{}0$ in the limit $L\tend\infty$ 
in the disordered PM phase. 

The magnetic susceptibility, $\chi{}_m$ and the heat capacity , $C_v$ are calculated from the magnetization and the energy 
fluctuations respectively
\equa{fluct}
  \chi_m = \frac{N}{T^*}\left[ \left(<m^2> - <m>^2 \right) \right] ~~~, ~~ C_v = \frac{1}{NT^{*2}} \left[ \left( <H^2> - <H>^2 \right) \right]
\auqe
Finally, in order to characterize the SG/PM transition, 
we use the standard spin-glass order parameter $q^2$ and the related spin-glass Binder cumulant
  \equa{sg_q}
  q^2 = \sum_{\al\be} |q_{\al\be}|^2 ~ ; ~~ 
  B_{sg} = \frac{1}{2}\left( 11 - 9 \frac{q_4}{q_2^2} \right)
\auqe
where $q_{\al\be}=\sum_i \mu_{i\al}^{(1)}\mu_{i\be}^{(2)}/N$, and the superscripts $(1)$ and $(2)$ denote two
independent replicas of an identical sample. 
We also use the spin-glass correlation length 
\equa{xi_sg}
  \xi = \frac{1}{sin(k_m)} \left( \frac{q_2(0))}{q_2(k_m)} -1 \right)^{1/2} ~~
  \textrm{with}  ~~~ q_2(k) = \sum q_{\al\be}(k) q^*_{\al\be}(k) 
  ~~~~ q_{\al\be}(k) =  \frac{1}{N} \sum_i \mu_{i\al}^{(1)}\mu_{i\be}^{(2)} e^{i\vec{k}\vec{r}_i}.
\auqe
%
 \section {Results}
 \label   {results}

\begin{figure} [h]
    \includegraphics [width = 0.80\textwidth , angle = -00.00]{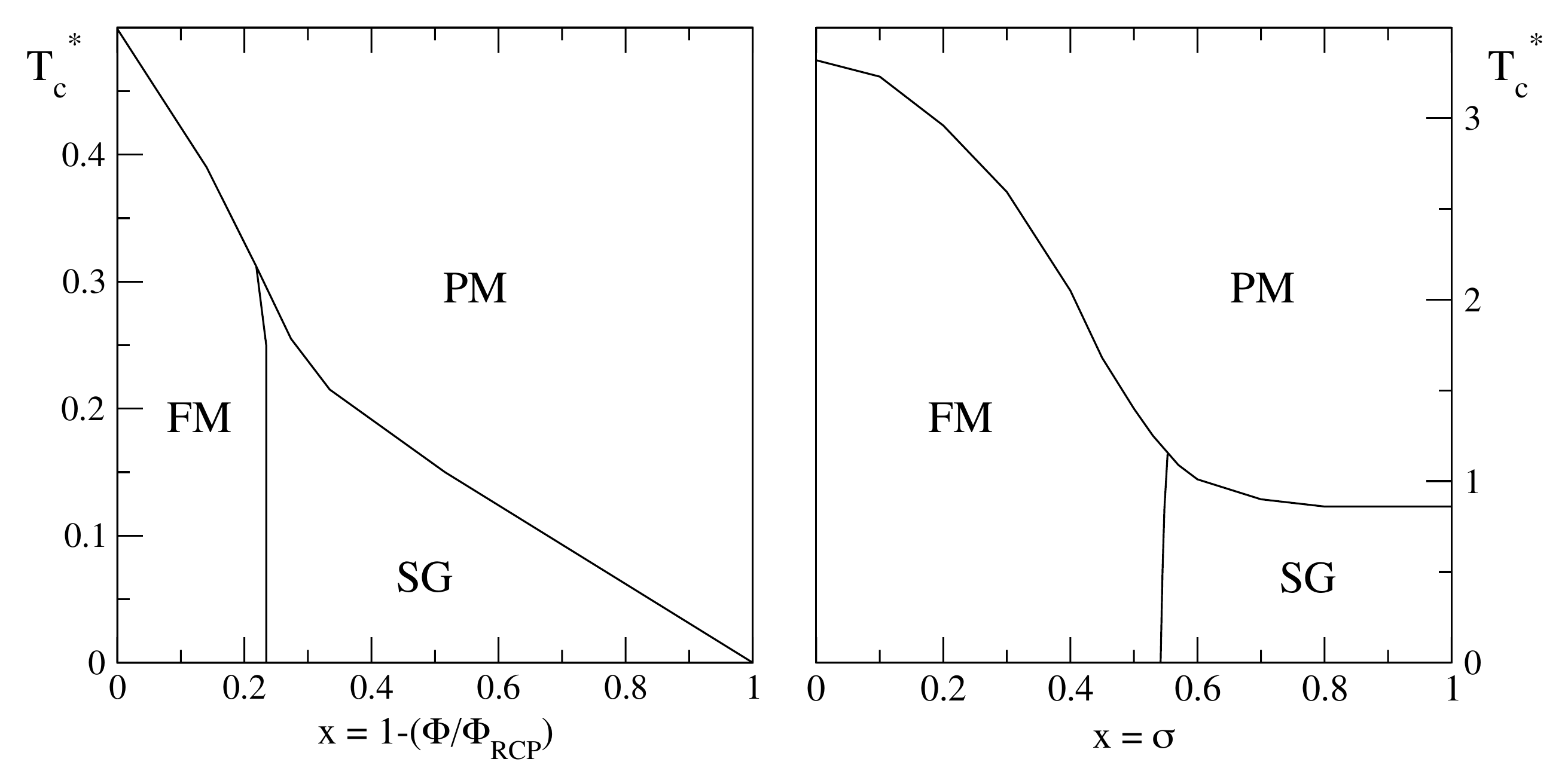}
      \vskip +0.00\textheight
    \caption {\label {schema}
    Schematic representation of the 3D isotropic frozen DHS with the HS like distribution in terms of the disorder 
    control parameter $x$ (see text). Left: model free of MAE. The FM/SG line is located at 
    $\Phi=\Phi_s\simeq{}0.49$. From Ref.~\cite{alonso_2020b}.
    Right: DIM with textured distribution of the Ising axes from Ref.~\cite{alonso_2019}.
    }
    \end{figure}

\begin {figure} [h] 
   \includegraphics [width = 0.95\textwidth , angle =  00.00]{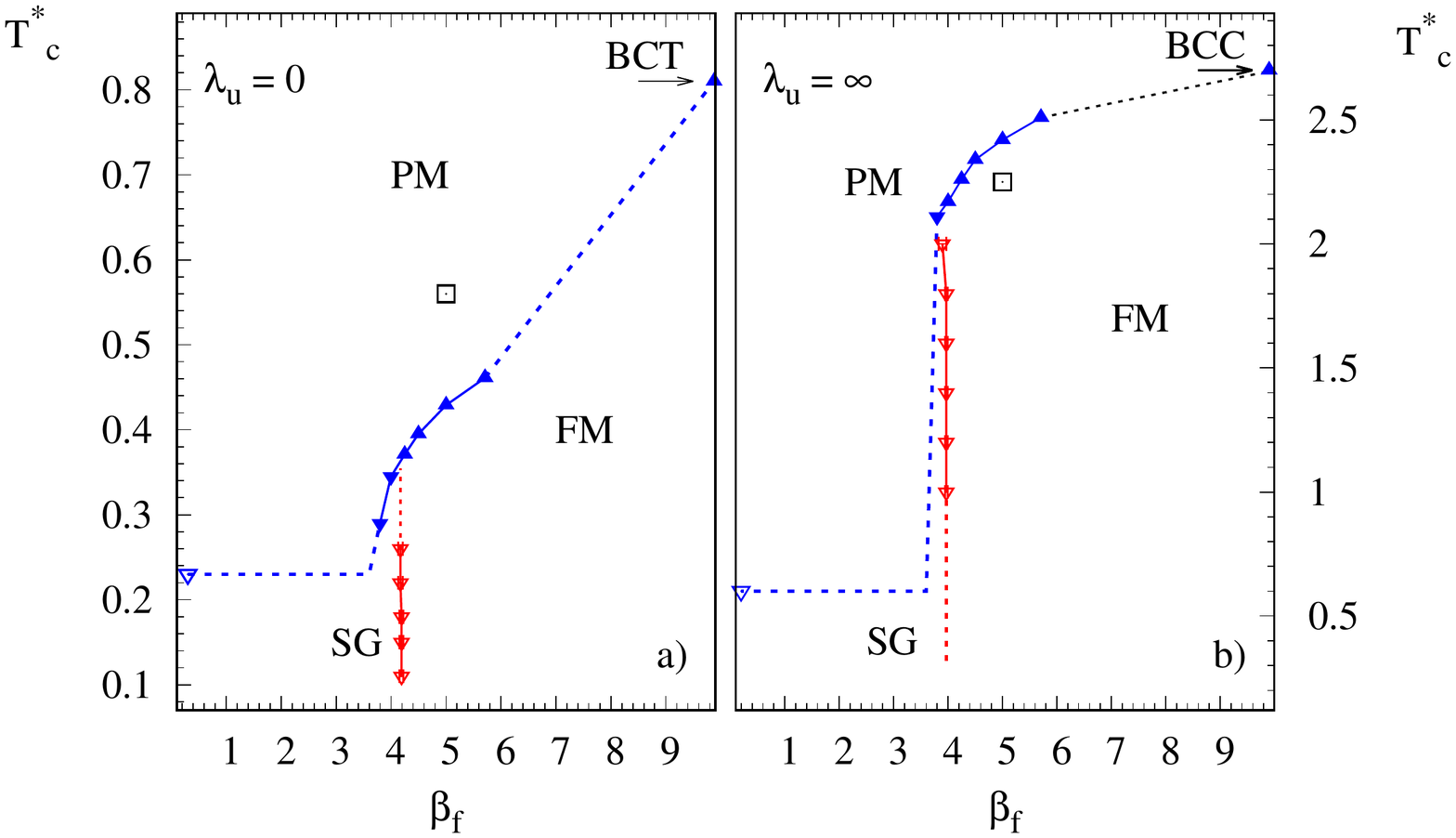}
   \vskip -0.03\textheight
   \caption {\label {diag_ph_dipfr_045}
   Phase diagram in the plane ($\be_f, T^*$) for $\Phi=0.45$.
   Open squares correspond to the PM/FM transition at $\Phi=0.262$ and $\be_f=8.5$ compared
   to the $\be_f=5$ case according to the $\la_s(\be_f)$ curve shown on figure~(\ref{dhs_45_lam_st}).
   a) Frozen dipolar hard sphere system. b) Frozen dipolar Ising model.
   Downward and upward triangles correspond to the PM/FM and either the SG/PM and the SG/FM lines. 
   The dotted lines are either continuations of the calculated PM/FM or SG/FM lines
   or an estimation of the SG/FM line. For the latter we assume a nearly isotropic distribution of the DHS 
   for $\be_f<3.7$ leading to the value of $T^*_c$ close to that obtained for the HS distribution (downward open triangle). 
   For the SG/FM we emphasize
   that no reentrance behavior can be deduced from our determination of the SG/FM lines for both
   the frozen DHS and the frozen DIM. 
   }
   \end {figure}

One of the main results of the paper is the magnetic phase diagram of the frozen DHS for $\Phi<\Phi_s$
in terms of the inverse freezing temperature $\be_f=1/T^*_f$, 
displayed either in the absence of MAE on figure~(\ref{diag_ph_dipfr_045}a), or in the strong MAE limit 
on figure~(\ref{diag_ph_dipfr_045}b), where the model becomes the dipolar Ising model.
Conversely to the examples shown in figure~(\ref{schema}), for convenience we present the phase diagram in terms of
$\be_f$ instead of $x$, the only difference being that the amount of disorder decrease with the increase of $\be_f$.
The disorder control parameter should be $x=T^*_f=1/\be_f$. 
The qualitative similitudes with the phase diagrams shown in figure~(\ref{schema}) is clear.
Here we find FM order for high values of $\be_f$ for which the structural anisotropy develops as evidenced by
the structural nematic order parameter,
$\la_s$, shown in figure~(\ref{dhs_45_lam_st}). This latter quantifies the structural anisotropy 
of the liquid DHS and is directly related to $\be_f$ since it is nothing but the value of the inverse temperature $\be=1/T^*$ 
at which the DHS structure is frozen to define the frozen DHS and frozen DIM models (see section~\ref{liq_dhs_045}). 
Notice that equivalently, the phase diagrams can be represented in the $(T^*_c,\la_s)$ plane.
We now discuss in more details the properties of the models and the way in which the phase diagrams are built.
 
  In the following, we present our results for the two volume fraction, $\Phi=0.45$ and 0.262.
  We have chosen $\Phi=0.45$ as the volume fraction upper bound since it is slightly smaller than 
the threshold value necessary to reach the FM ordered phase on an isotropic hard sphere like 
frozen structure~\cite{alonso_2020b} (see section~\ref{dipfr_45}). 
In order to test the persistence of our findings, 
we have considered a lower density case, $\Phi=0.262$ ($\rho^*=6\Phi/\pi=0.5$). 
It corresponds to a dilute liquid state and it is sufficiently far from the volume fraction
range where the DHS presents a self assembly behavior and undergoes structural 
transitions~\cite{Kantorovich_2015}. 
Indeed, from Ref.~\cite{camp_2000} and the extrapolated phase diagram given in Ref.~\cite{holm_2005} we 
can estimate that the DHS behaves as a bulky liquid only beyond $\Phi=0.18$ ($\rho^*=0.35$). 

 \subsection {Liquid DHS}  
 \label {liq_dhs_045}
\begin {figure} [h]  
   \includegraphics [width = 0.80\textwidth , angle =  00.00]{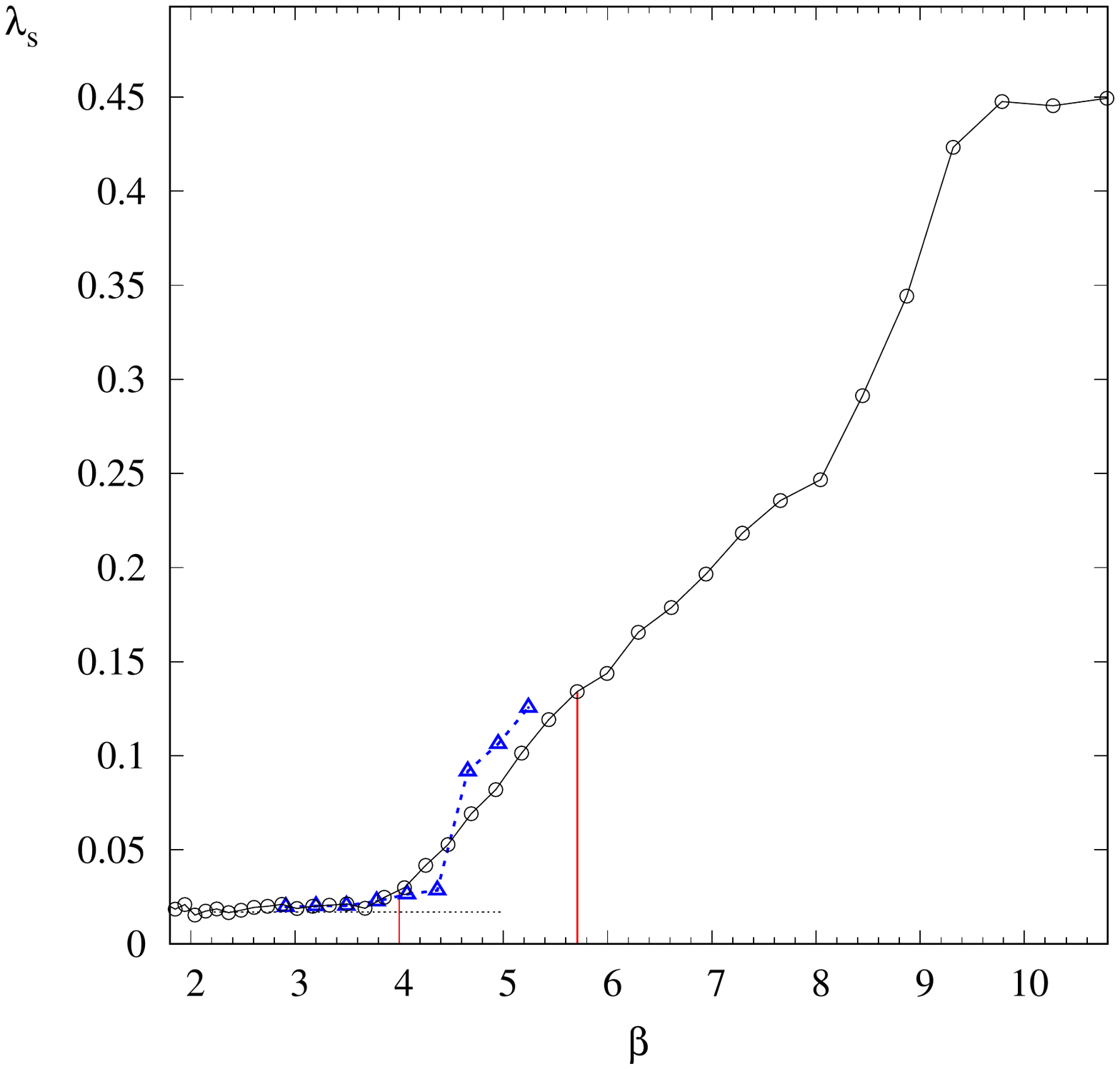}
   \vskip -0.03\textheight
   \caption {\label {dhs_45_lam_st}
   Structural nematic order parameter, $\la_{s}$ relative to the set of nearest neighbors bonds in the DHS 
   fluid at $\Phi=0.45$ in terms of the inverse temperature $\be=\ep_d/k_{B}T$ 
   calculated on the $N=1177$ system size. The vertical lines delimit the range of inverse temperatures chosen 
   for the frozen structures. Dashed line /triangles is the result of $\la_{s}$ at $\Phi=0.262$ and $N=728$ displayed in
   terms of $(r\be)$ where $r$ is the ratio of volume fractions ($r=0.262/0.45$) to take into account
   the $\be$ dependence through $\be\Phi$ at low structural anisotropy.
   }
   \end {figure}

\begin {figure} [h]  
   \includegraphics [width = 0.80\textwidth , angle = -00.00]{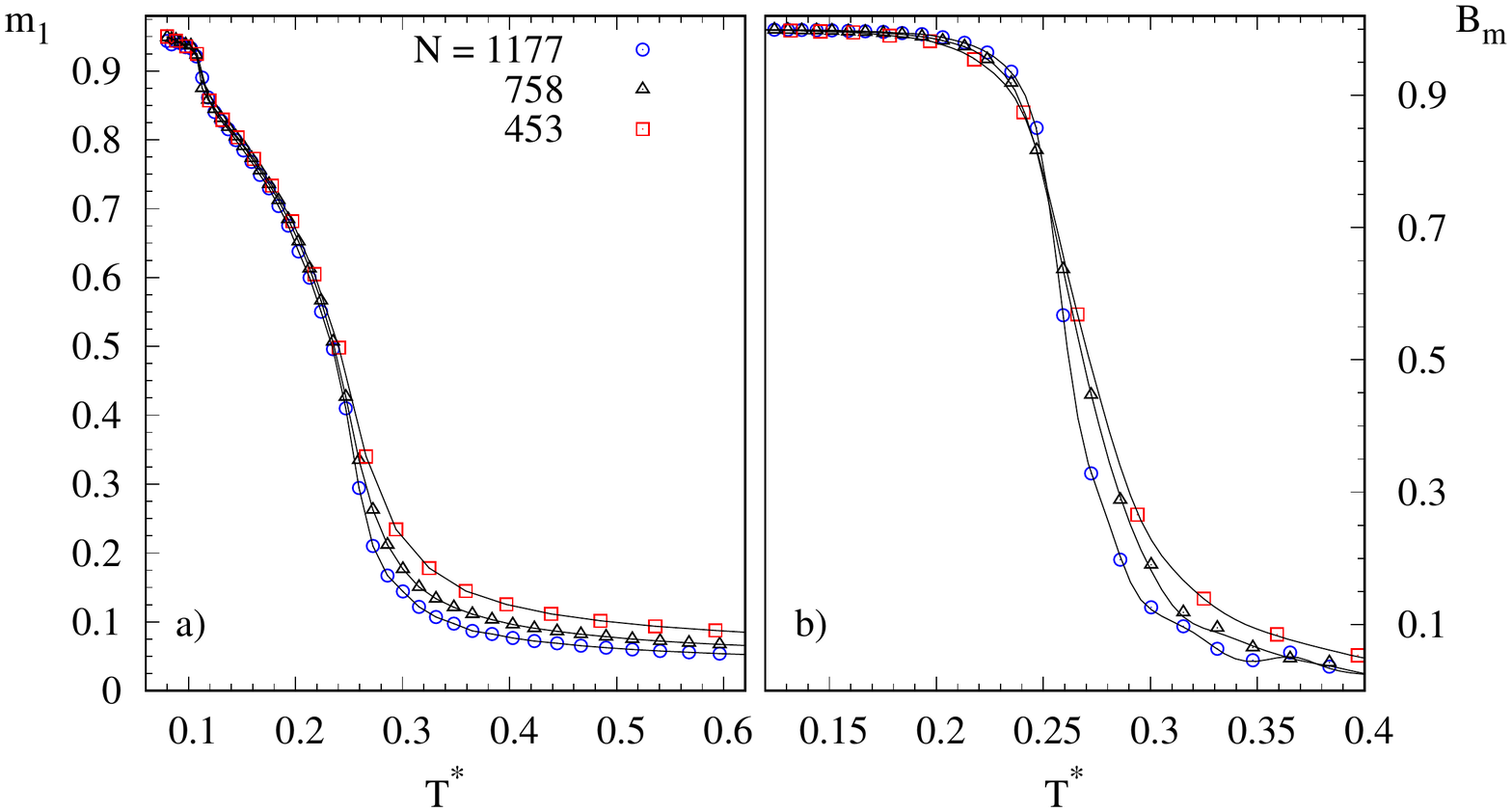}
   \vskip -0.03\textheight
   \caption {\label {dhs_45_m_bm}
   a) Magnetization and b) Binder cumulant $B_m$ of the DHS at $\Phi=0.45$
   and sizes ranging from $N=453$ to 1177 in terms of $T^*=k_{B}T/\ep_d$.
   The lines are obtained from the reweighting method.
      }
   \end {figure}

\begin {figure} [h]  
   \hskip -0.03\textwidth
   \includegraphics [width = 0.99\textwidth , angle =  00.00]{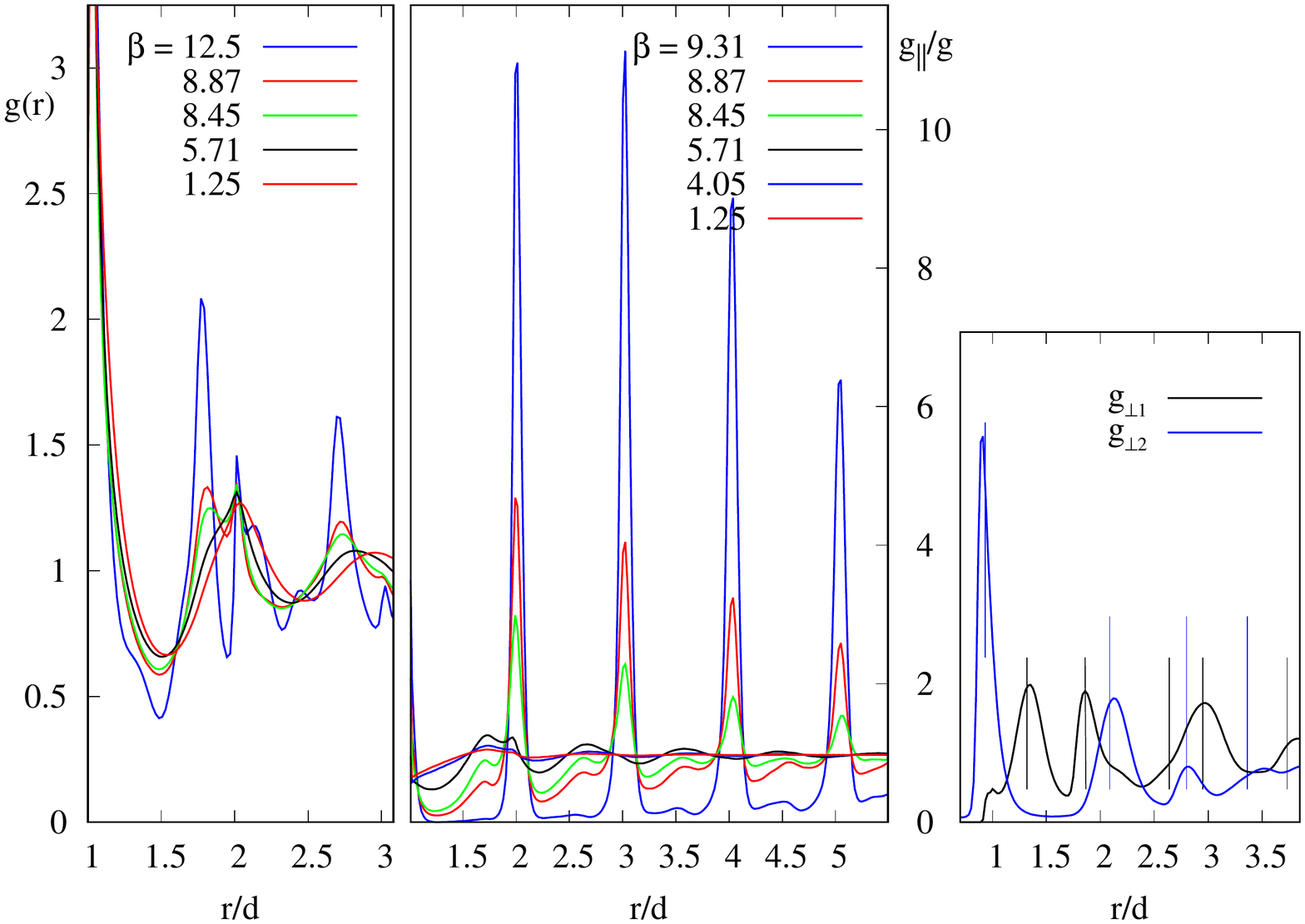}
   \vskip -0.03\textheight
   \caption {\label {dhs_45_anis}
   Left : Pair distribution function $g(r)$ of the DHS fluid at $\Phi=0.45$ and different values of the 
   inverse reduced temperature $\be=1/T^*$ as indicated, simulated with $N=1177$. The $g(r)$ for
   $\be\leq{4.05}$, corresponding to the PM/FM transition are very close to each other.
   Center : 
   Anisotropy of the DHS structure displayed from the ratio $g_{\parallel}(r)/g(r)$ for different values 
   of the inverse temperatures. The $\be=1.25$ and $4.05$ cases are nearly indistinguishable. The onset of 
   the BCT like solid structure with particles aligned at contact along the $c-$ axis beyond $\be=8.87$ is
   clearly seen.
   Right : $g_{\perp{1}}(r)$ and $g_{\perp{2}}(r)$ corresponding to particles in the basal and median planes of the 
   expected BCT structure at very low temperature (see text and note~\cite{note_bct}) for $\be=12.5$. 
   The vertical lines denote the locations of the BCT nodes mentioned in note~\cite{note_bct} for $a/c=1.32$.
   }
   \end {figure}

Since in the following subsections we use frozen distributions of particles resulting from the liquid DHS,
we first present here the necessary results of our own simulations on the liquid DHS.
Simulations on this system have already been reported in the 
literature~\cite{wei_1992a,weis_1993,weis_2006,levesque_2014}. Here,
on the first hand we bring a new point in the FM/PM transition line at
low density $\phi=0.262$, and on the other hand we clarify
the onset of the structural anisotropy related to the PM/FM transition.
\\ \indent
 In figure~(\ref{dhs_45_m_bm}a) we show the magnetization in terms of $T$ of the DHS 
for 3 system sizes, namely $N=1177$, 758 and 453. 
Two important features can be deduced. First $m(T)$ becomes size independent at $T$ lower than 
a threshold value of $\sim{}0.2$ indicating a FM order with a finite $m_1(T\tend{}0)$ limit.
Second, $m_1$ 
presents a clear jump for the 3 system sizes at $T^*\simeq{}0.12$ which is interpreted as the 
onset of the structural liquid/solid transition where the tetragonal structure builds up.
The PM/FM transition temperature is obtained from the crossing point of the magnetization Binder
cumulant curves $B_m(T,N)$ (see figure~(\ref{dhs_45_m_bm}b)) and found at $T^*_c=0.25\pm{}0.015$.
This result is in agreement with that of Ref.~\cite{weis_2006} corresponding to 
the reduced density $\rho^*=0.88$ ($\Phi=0.4607$). Notice that as in~\cite{weis_2006} 
we also get a scaling behavior 
of $B_m$ and $m_2$ with the critical exponents $\nu\simeq{}0.709$ and $\be/\nu\simeq{}0.525$, 
coherent with the 3D Heisenberg universality class although our
system sizes are too small for an efficient determination of the critical exponents.
The DDI induced anisotropy in the DHS at temperatures lower than $T^*_c$ is quantified by a
structural nematic order parameter $\la_s$~\cite{gingras_1995}, characterizing the next neighbor 
bonds distribution which must not be confused with the nematic order parameter $P_2$ defined 
from the moment distribution (see section~\ref{obs}). 
$\la_s$ is defined as the largest eigenvalue of the nematic tensor
$\bar{Q}_{s}=(1/N_{nn})\sum_{nn} (3\hat{l}_{nn}\hat{l}_{nn}-\bar{I})/2$, where 
$\{\hat{l}_{nn}\}$ is the set of next neighbors bonds in the system, $N_{nn}$ their total
number and $\bar{I}$ the identity tensor. 
The result for $\la_{s}$ is shown on figure~(\ref{dhs_45_lam_st}) in the case $N=1177$.
We clearly see that the structure of the DHS remains isotropic ($\la_s\simeq{0}$) up to
$T^*\simeq{0.27}$ ($\be=1/T^*\simeq{3.70}$) close to the PM/FM transition temperature, 
$T^*\simeq{0.25}$, then increases strongly up to $T^*\sim{0.12}$ ($\be\sim{8.5}$) where the 
jump in $m(T^*)$ is observed and is interpreted as the onset of the crystallization of 
the DHS. This picture corroborates that given in Ref.~\cite{wei_1992a}. 
More precisely the present simulation
shows that the anisotropy in the fluid starts to set up in the vicinity of the PM/FM
transition temperature and strengthens with the decrease of $T$ up to the onset of the 
solid structure.
Another way to characterize the anisotropy of the fluid at $T^*<T^*_c$ comes from the pair distribution
function, $g(r)$. On the one hand we compare $g(r)$ to its longitudinal component $g_{\parallel}(r)$
limited to the direction parallel to nematic direction $\hat{d}$ and on the other hand we introduce the
two transverse components, $g_{\perp{1}}(r)$ and $g_{\perp{2}}(r)$ corresponding to particles $(i,j)$ 
in the plane normal to $\hat{d}$ and with 
$\vec{r}_{ij}\ldotp\hat{d}$ or $(\vec{r}_{ij}\ldotp\hat{d}-1/2)\in[-\gD,\gD]$
with $r_{ij}$ in unit of $d$ respectively. The thickness $\gD$ must be smaller than 0.5 for coherence
and is chosen as $\gD=0.4$.
The comparison of $g$ with $g_{\parallel}$ leads to a direct estimation of the structural anisotropy,
see the center panel of figure~(\ref{dhs_45_anis}),
and the analysis of the peak positions of $g_{\perp{1}}(r)$ and $g_{\perp{2}}(r)$ shows that the
DHS at low temperature orders according to the BCT structure with particles at contact along the 
nematic direction $\hat{d}$, namely $c=1$~\cite{note_bct}, 
and close to the result obtained by Levesque and Weis~\cite{levesque_2014} at a similar density
(see the right panel of figure~(\ref{dhs_45_anis})).
We have to note that for the lowest temperature used in the present work, the structure is not a perfect BCT
lattice at least because of the cubic shaped simulation box, and presents defects leading to a 
value for $a/c\simeq1.32$ instead of 1.525, the value for the perfect BCT lattice at $\Phi=0.45$.
The same behavior was obtained by Levesque and Weis~\cite{levesque_2014} in their simulation including 4\;000
particles. \\

\begin {figure} [h] 
   \includegraphics [width = 0.80\textwidth , angle =  00.00]{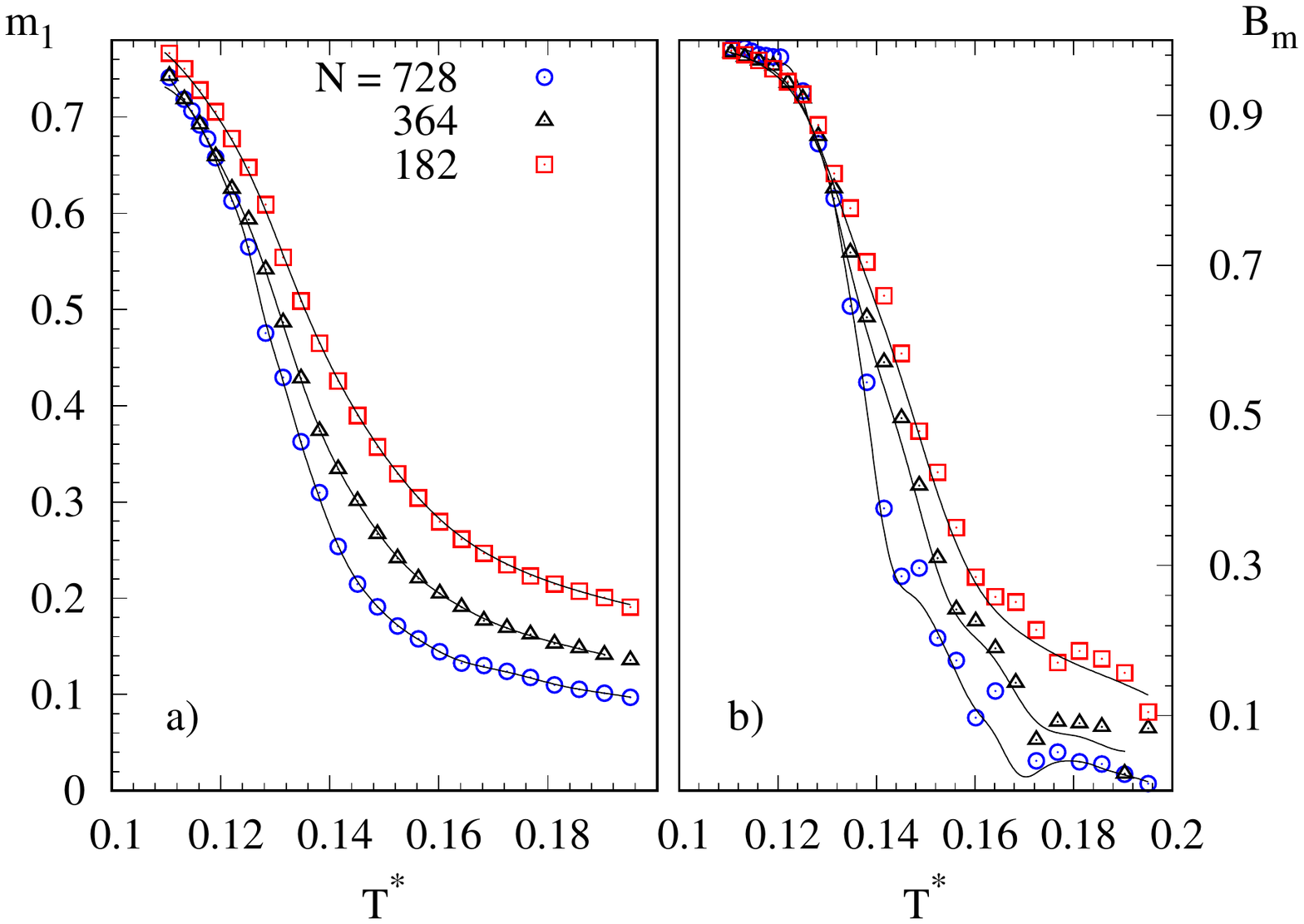}
   \vskip -0.03\textheight
   \caption {\label {m_bm_dhs_026}
   DHS model at $\Phi=0.262$. a) Magnetization and b) magnetization Binder cumulant 
   in terms of $T^*$ for systems sizes ranging from $N=182$ to 728.
   The lines are obtained from the reweighting method.
   }
   \end {figure}

 At low density, the simulations are more intricate due to the larger flexibility in the formation of local structures
 (chains, rings). 
 The first consequence is the absence of the onset of large scale solid structure, at least in the
 temperature range investigated here ($T^*\ge{}0.10$)~\cite{note_tphi} for $\Phi=0.262$ (i.e. $\rho^*=0.5$),
 considered as a low density characteristic case.
 The DHS presents a magnetization 
 (see figure~(\ref{m_bm_dhs_026}a)) 
 with the same features as that obtained at $\Phi=0.45$ with a 
 nearly independence in terms of the system size $L$, for $T^*$ smaller than $c.a.~0.12$. 
 From the crossing point of the magnetization Binder cumulant curves in terms of $T^*$ 
for different system sizes ($N=$~182, 364 and 728) 
shown on figure~(\ref{m_bm_dhs_026}b)
we get a PM/FM transition at $T^*_c{}=0.130\pm{}0.005$. 

\begin {figure} [h] 
   \includegraphics [width = 0.80\textwidth , angle =  00.00]{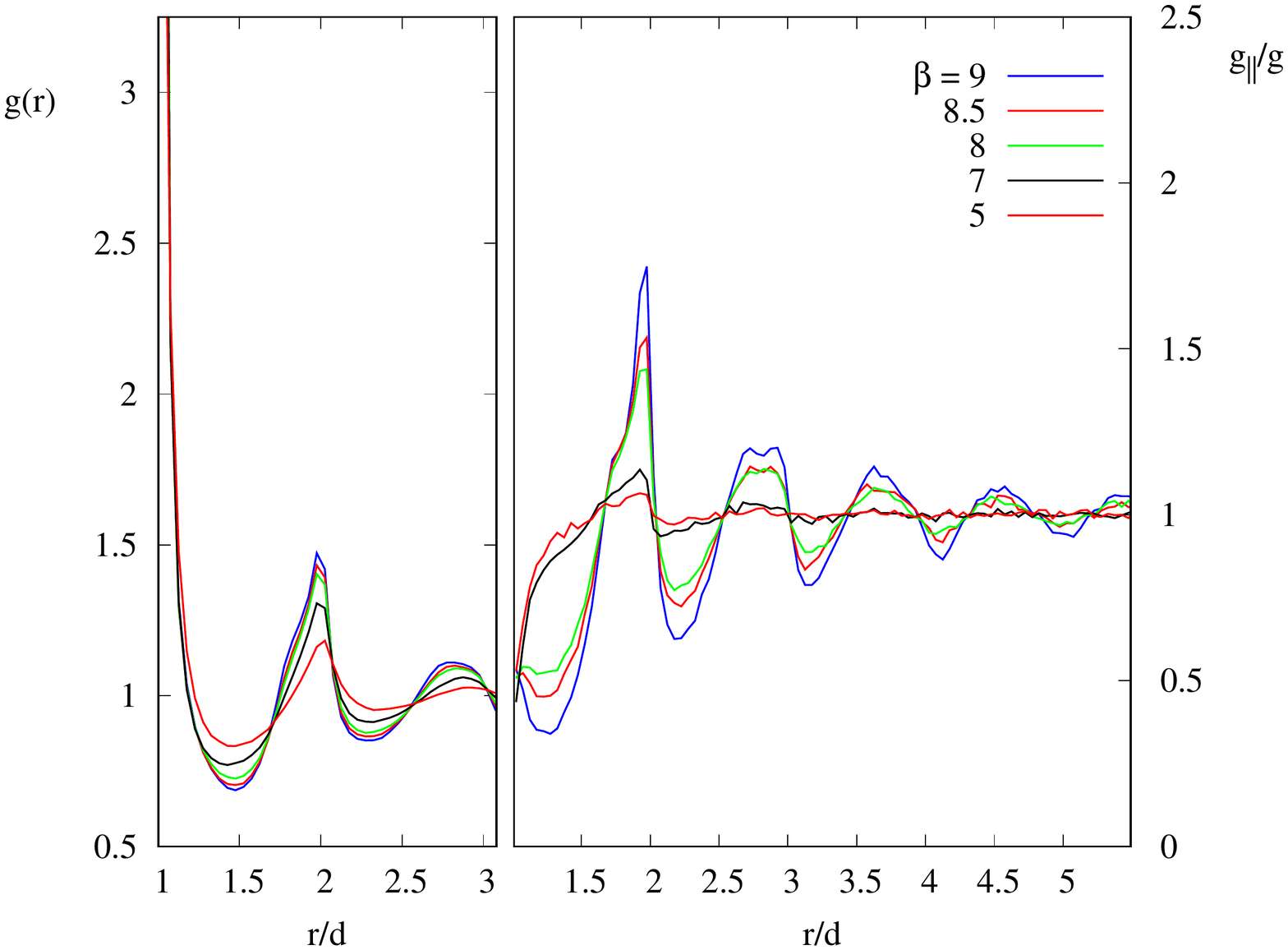}
   \vskip -0.03\textheight
   \caption {\label {struct_dhs_026}
   Pair distribution functions, $g(r)$ and $g_{\parallel}(r)/g(r)$ of the DHS model at $\Phi=0.262$
   computed with $N=728$.}
   \end {figure}

 As is the case for $\Phi=0.45$, the onset of structural anisotropy together with its 
 increase with the decrease of $T^*$ is evidenced both from the structural nematic order parameter $\la_s$ and 
the pair distribution function, $g(r)$ and its longitudinal component, $g_{\parallel}(r)$.
The evolution with $T^*$ of the ratio $g_{\parallel}(r)/g(r)$ (see figure~\ref{struct_dhs_026}),
is qualitatively similar to that obtained for $\Phi=0.45$, excepted the onset of the ordered
phase obtained in the latter case at low $T^*$.
 Here, the growing structure of the peak at $r/d\sim{2}$ with the decrease of $T^*$ is associated to the head to
 tail formation of chain dipoles. 
It is worthwhile to note that the
onset of the structural anisotropy occurs at a value of $T^*$ only slightly smaller than $T^*_c$ and moreover
that in the vicinity of $T^*_c$, $\la_s$ presents a similar increasing rate in terms of $\Phi{}T^*$ for 
$\Phi=0.262$ and 0.45 (see figure~(\ref{dhs_45_lam_st})), in agreement with the note~\cite{note_tphi}.

 \subsection {Dipolar system with frozen distribution}
 \label {dipfr}
  Once the structure of the DHS in terms of $T^*$ is determined, we consider the model of dipolar 
hard spheres with frozen distributions of the particles being obtained as the equilibrium configurations 
of the DHS at conveniently chosen freezing temperatures. Thus we consider a set of configurations obtained 
on the DHS at a fixed inverse freezing temperature, say $\be_f$ ($\be_f>1/T^*_c(DHS,\Phi)$), 
as a set of realizations for the distribution of dipoles of the frozen DHS model. 
All these configurations are characterized by a uniaxial broken symmetry and by the same structural 
anisotropy quantified by $\la_s(\be_f)$. 
In the absence of MAE, $\la_u=0$, the frozen DHS model built in this way 
is fully parametrized by $\be_f$ (or equivalently $\la_s$). 
Conversely the structure in the frozen DIM, in the infinitely large MAE coupling limit $\la_u\tend\infty$, 
includes both the particles and the easy axes distributions. We consider 
that in the liquid the $\{\hat{n}_i\}$ follow instantaneously the $\{\hat{\mu}_i\}$ ($i.e.$ vanishingly
small Brownian relaxation time). Therefore, in the frozen structure, we set 
$\hat{n}_i=\hat{\mu}_i$, the $\{\hat{\mu}_i\}(\be_f,\Phi)$ being the equilibrium
configuration of the moments in the DHS at $T^*=1/\be_f$ and $\Phi$. Accordingly, in addition to the structural
anisotropy, quantified by $\la_s$, we have a texturation of the $\{\hat{n}_i\}$ distribution, quantified 
by the nematic order parameter $P_2$ introduced in section~\ref{obs}. Of course, since both $\la_s$ and $P_2$ 
are determined by $\be_f$ and $\Phi$ they are not independent parameters, and the model is still 
parametrized by $\be_f$.

In the following we present the phase diagram of the frozen DHS model ($\la_u=0$)
and of its frozen DIM limit ($\la_u\tend\infty$). 

 \subsubsection {Frozen dipolar hard sphere model free of MAE. $\Phi=0.45$}
 \label {dipfr_45}
\begin {figure} [h]  
   \includegraphics [width = 0.90\textwidth , angle = -00.00]{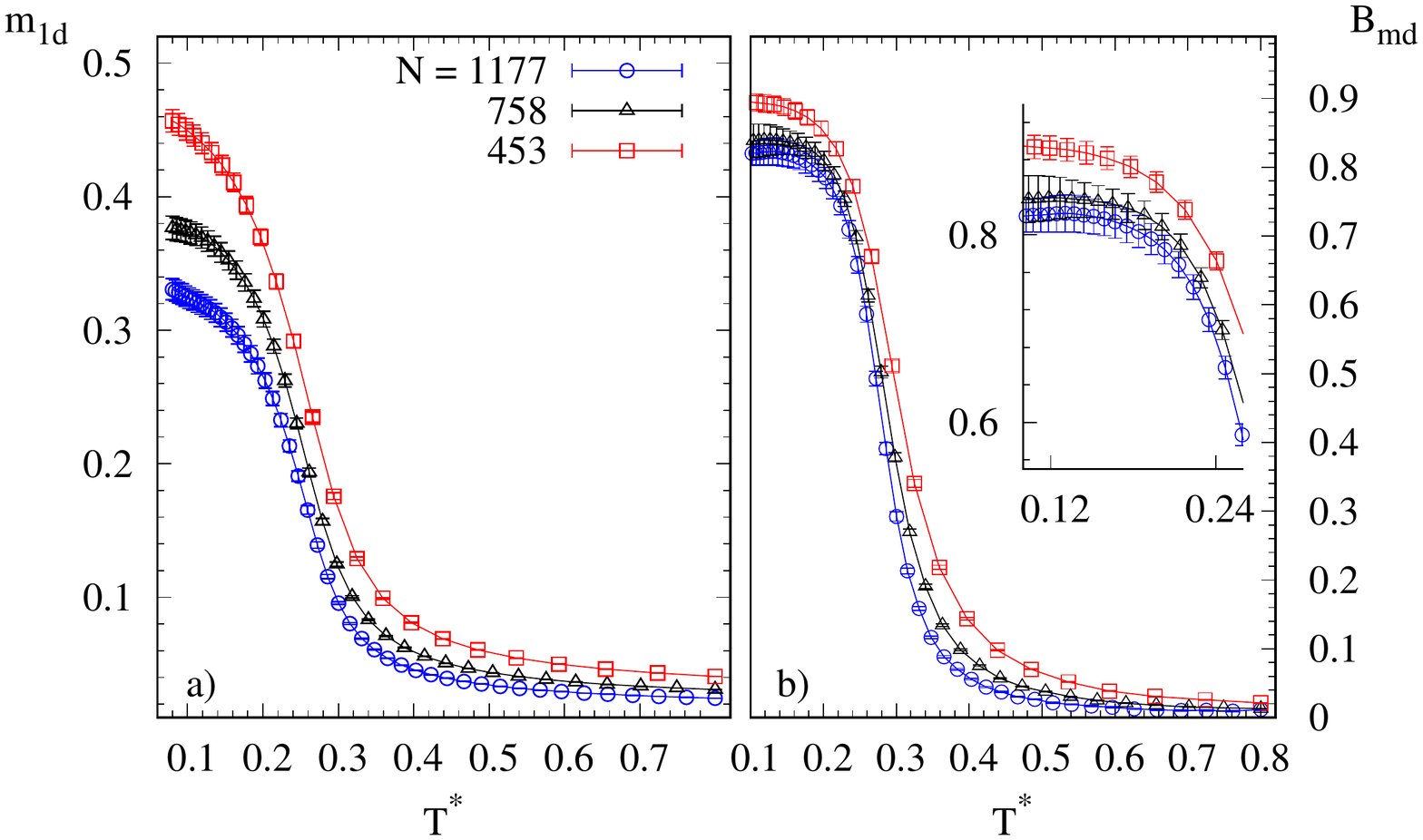}
   \vskip -0.03\textheight
   \caption {\label {bm_hs_45}
   a) Magnetization, $m_{1d}$ and b) Binder cumulant $B_{md}$ for the frozen dipolar model with the 
      HS structure at $\Phi=0.45$ and sizes ranging from $N=453$ to 1177 in terms of $T^*=k_{B}T/\ep_d$.
      }
   \end {figure}

\begin {figure} [h]  
   \includegraphics [width = 0.90\textwidth , angle = -00.00]{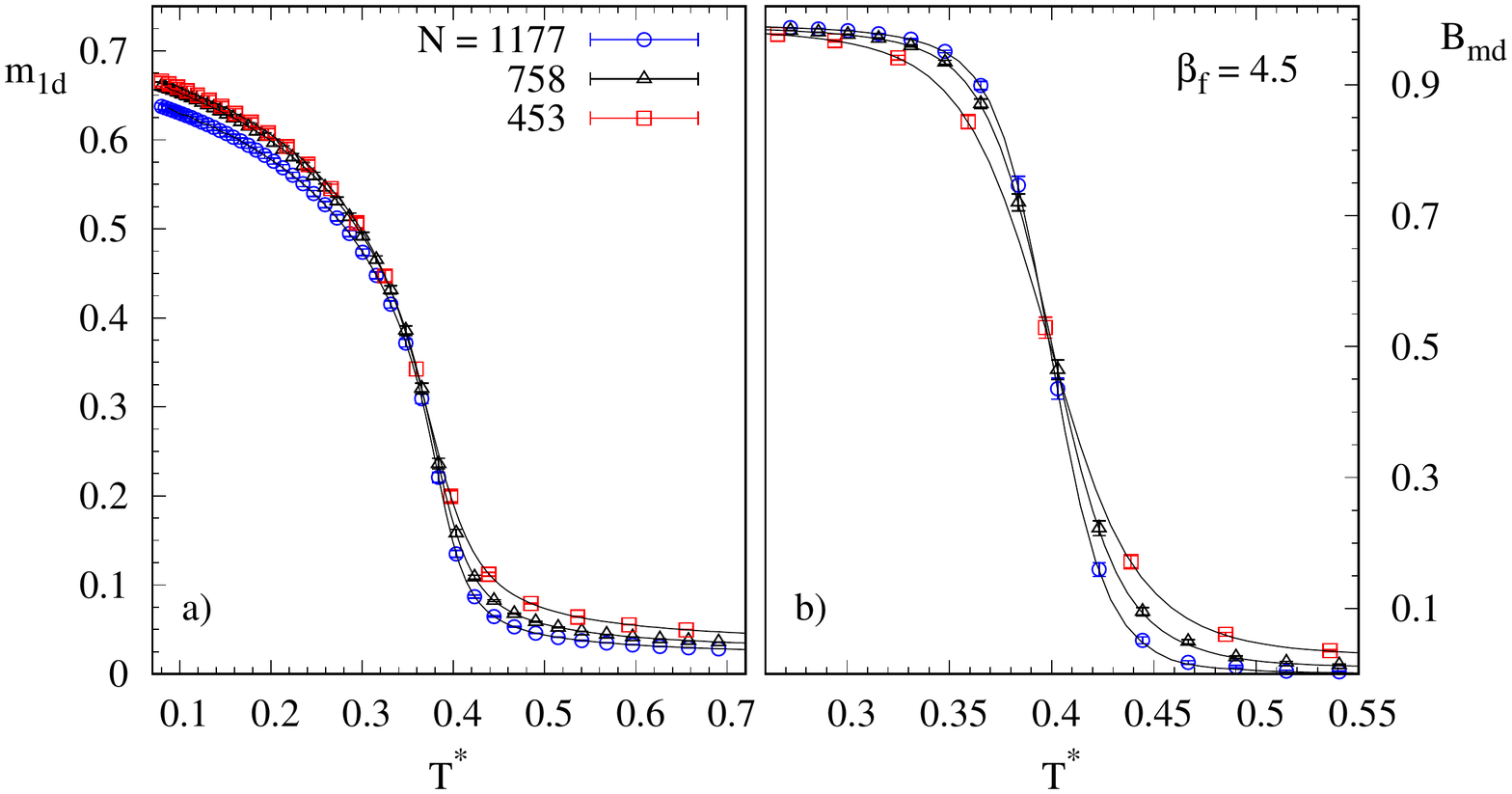}
   \vskip -0.03\textheight
   \caption {\label {dipfr_045_be45_m_bm}
   Frozen DHS model at $\be_f=4.5$ and $\Phi=0.45$. a) Magnetization $m_{1d}$ and b) Binder cumulant $B_{md}$ in terms of $T^*$.
    }
   \end {figure}

\begin {figure} [h]  
   \includegraphics [width = 0.92\textwidth , angle = -00.00]{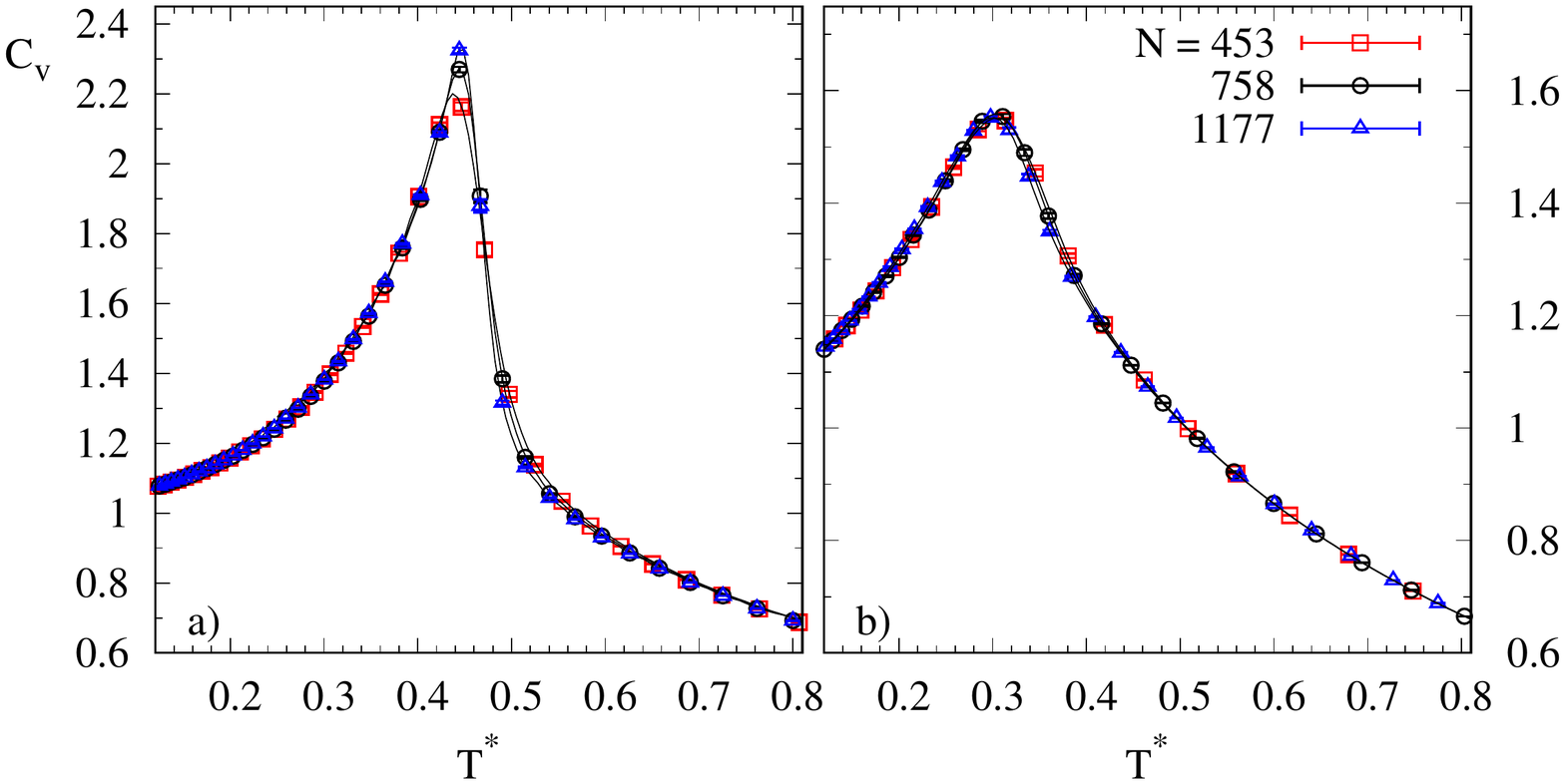}
   \vskip -0.03\textheight
   \caption {\label {dipfr_045_cv_be571-380}
   Heat capacity $C_v$ of the frozen DHS model at $\Phi=0.45$ and different systems sizes.
   a) $\be_f=5.71$; b) $\be_f=3.80$. The lambda like shape and the finite size effect in the vicinity of the maximum,
   features of the PM/FM transition clearly disappear at $\be_f=3.80$ where the system orders in a SG phase.
    }
   \end{figure}

\begin {figure} [h] 
   \hskip -0.20\textwidth
   \includegraphics [width = 0.75\textwidth , angle =  00.00]{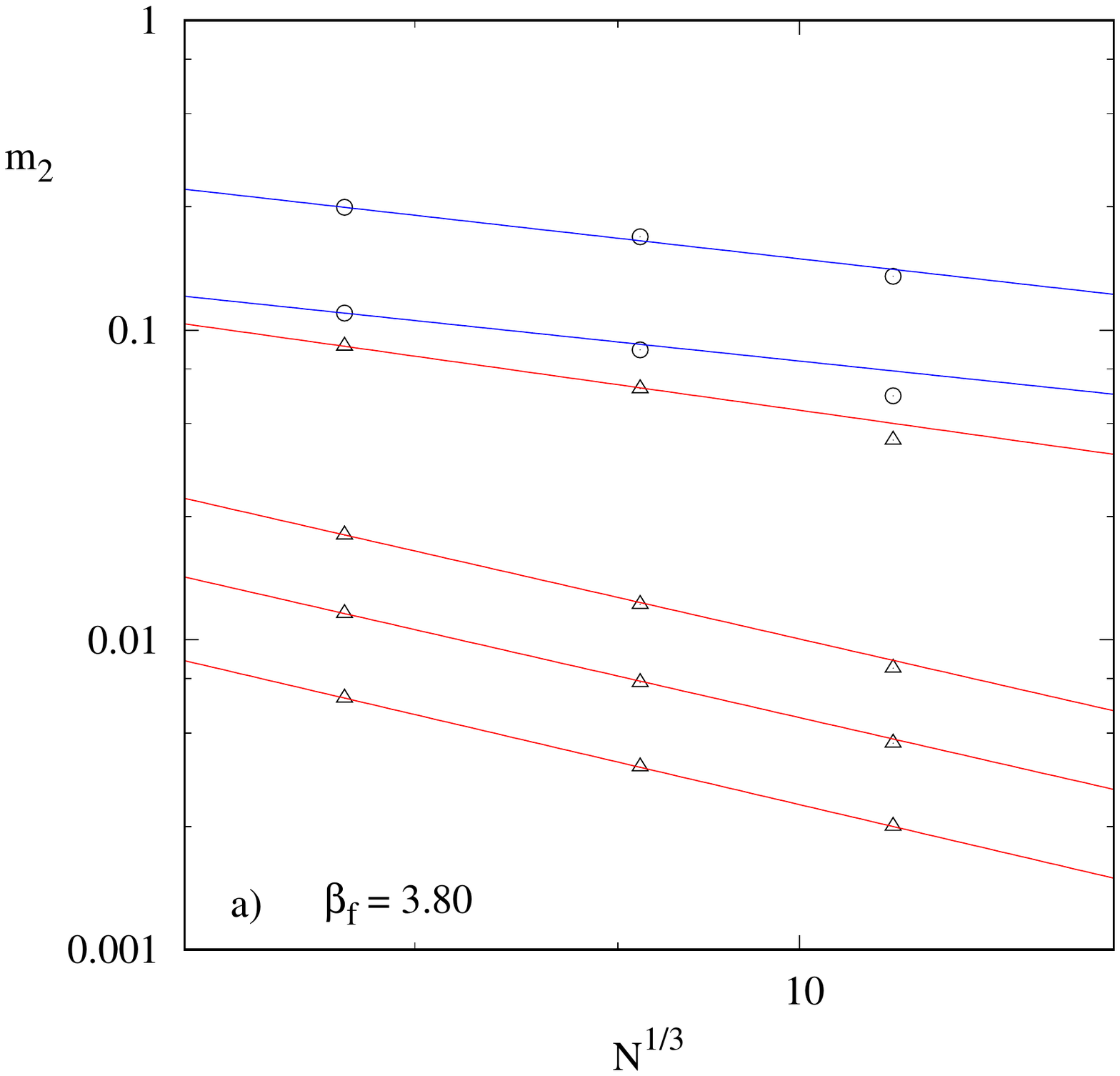}
   \hskip -0.31\textwidth
   \includegraphics [width = 0.75\textwidth , angle =  00.00]{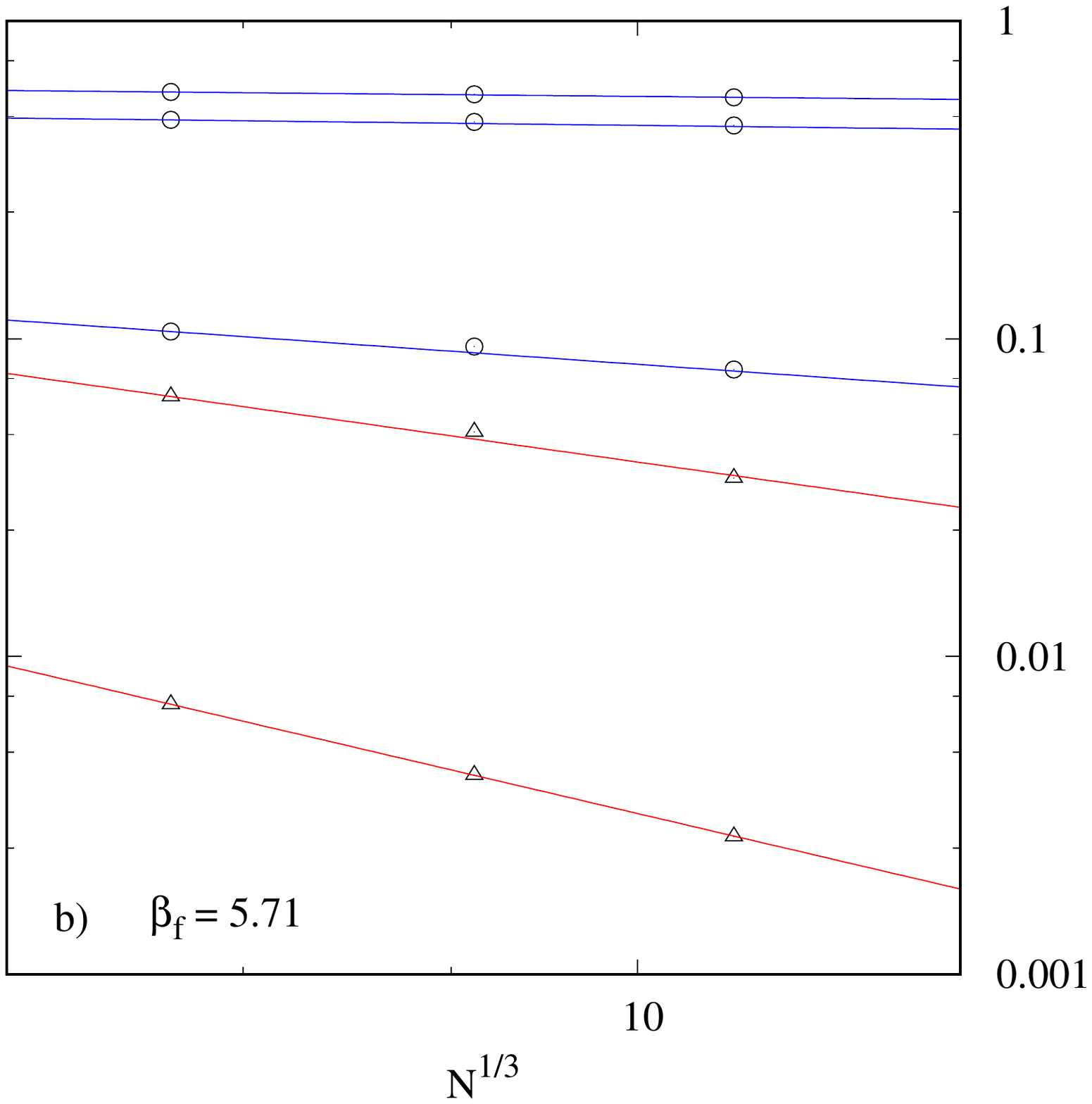}
   \caption {\label {lnm2_lnl}
   Moment $m_2$ in terms on $L=N^{1/3}$ in log scale for the frozen DHS model at $\Phi=0.45$.
   Open circles (triangles) and blue (red) lines correspond to $T^*<T^*_c$ ($T^*>T^*_c$).
   a) $\be_f=3.80$ and $T^*=0.10$, 0.28, 0.30, 0.40, 0.50 and 0.80 from top to bottom.
   b) $\be_f=5.71$ and $T^*=0.102$, 0.250, 0.454, 0.471 and 0.80 from top to bottom.
   a) and b) correspond to the SG and FM regions of the phase diagram since 
   in a) we get a decrease of $m_2$ with $N$ at all temperatures, and expect $m_2\tend{0}$
   at the thermodynamic limit,
   while in b) we get a vanishing slope of $m_2$ with $N$ at the lowest temperatures. 
   In both cases, the slope of the bottom line ($T^*=0.8$) is : $ln(m_2)\propto{}-3\;ln(N^{1/3})$.
   }
   \end {figure}

\begin {figure} [h] 
   \hskip -0.20\textwidth
   \includegraphics [width = 0.72\textwidth , angle =  00.00]{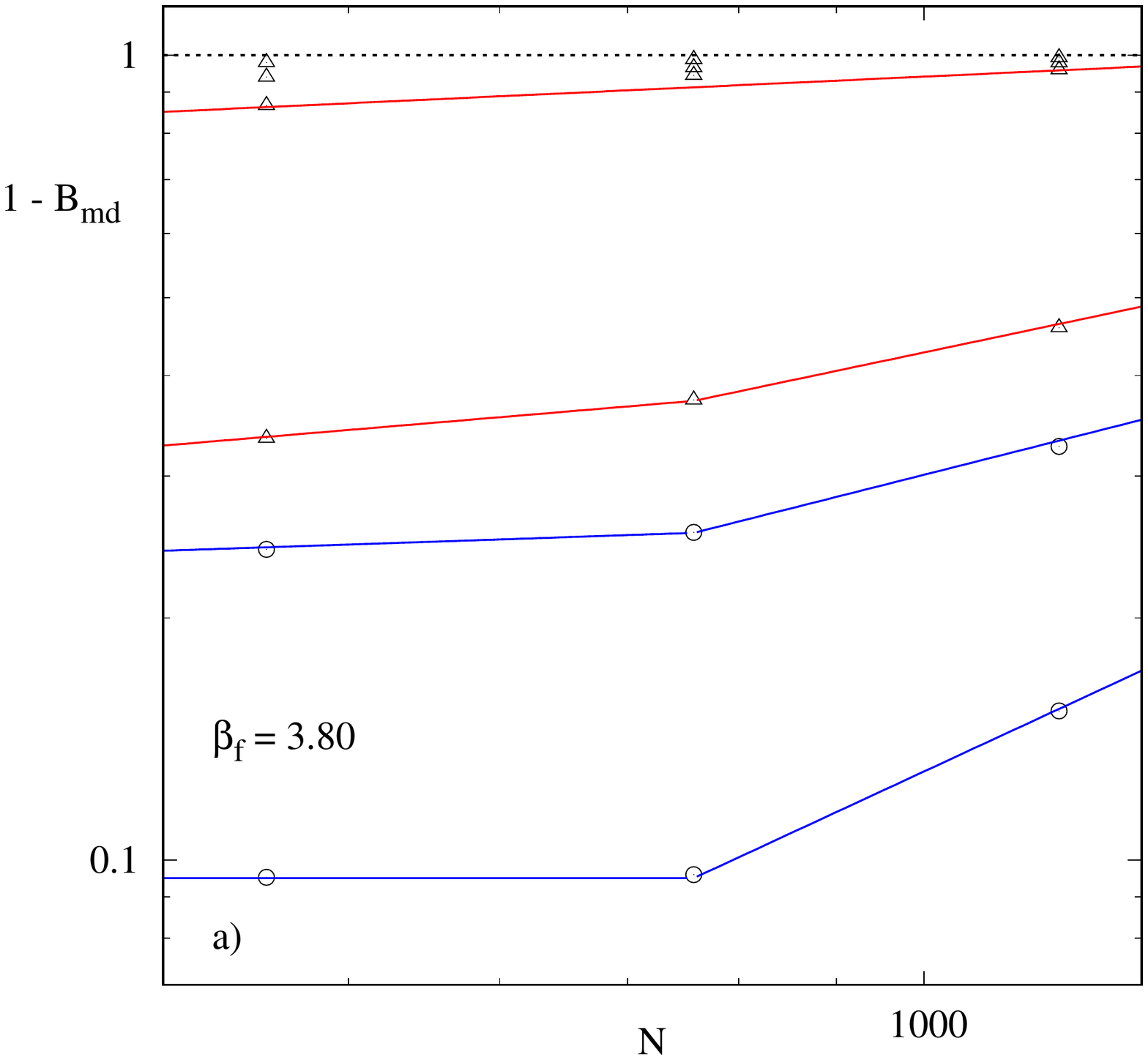}
    \hskip -0.28\textwidth
   \includegraphics [width = 0.72\textwidth , angle =  00.00]{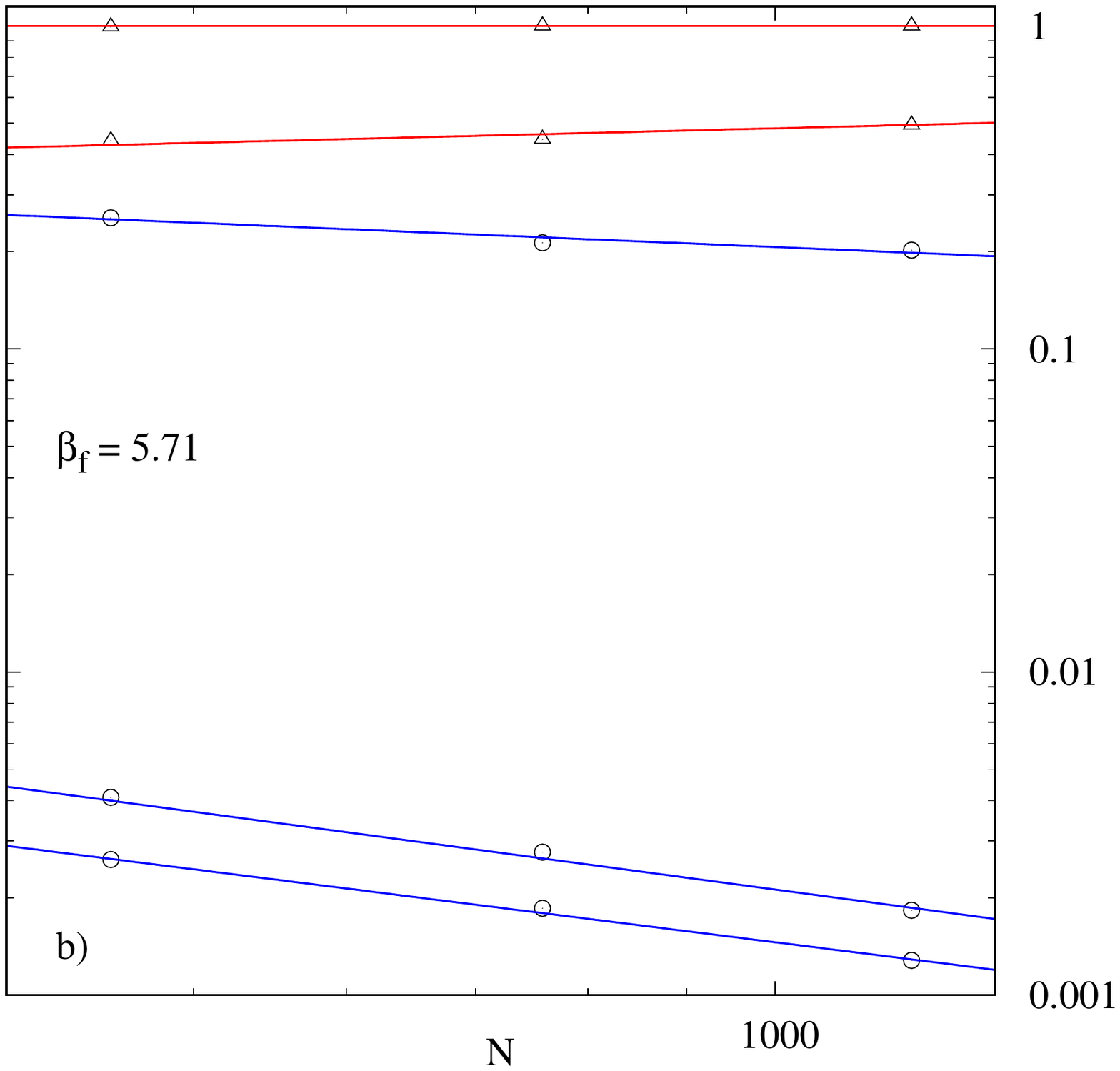}
   \caption {\label {1_moins_bm}
   Low temperature behavior of the magnetization Binder cumulant through $(1-B_{md})$ in terms of $N$
   for the frozen DHS model at $\Phi=0.45$. Open circles (triangles) and blue (red) lines correpond to
   $T^*<T^*_c$ ($T^*>T^*_c$).
   a) $\be_f=3.80$ and $T^*=0.10$, 0.28, 0.30, 0.40, 0.50 and 0.80 from bottom to top.
   b) $\be_f=5.71$ and $T^*=0.102$, 0.250, 0.454, 0.471 and 0.80 from bottom to top.
   The lines are linear interpolations of $ln(1-B_{md})$ in terms of $ln(N)$. 
   Notice the difference of the ordinate scales. 
   In a) we clearly see the behavior of the SG phase with increasing $(1-B_{md})$ with $N$ at $T^*<T^*_c$ while in b) the
   opposite is obtained.  
   }
   \end {figure}

 The first limiting case of the model is the high freezing temperature limit $\be_f\ll{}1/T^*_c(DHS)$ limit where 
the structure coincides with the pure hard sphere (HS) one since it does not depend on the DDI and is thus
isotropic. In this case, as expected~\cite{ayton_1997,alonso_2020b}, the system does not present any FM order 
at low temperature as can be deduced from the low temperature behavior of the magnetization $m_1$ and the absence 
of crossing point in the $B_m(T^*,N)$ curves, displayed on figure~(\ref{bm_hs_45}). 
Conversely, from the spin-glass Binder cumulant $B_{sg}$ and of the spin-glass correlation length $\xi/L$ 
a SG transition is evidenced at $T^*\simeq{}0.23\pm{}0.015$ in agreement with our preceding result~\cite{alonso_2020b}.  
 Then we consider the perfect lattice BCT with $c=d$ and $a/c=\sqrt{\pi/(3\Phi)}$  ($a/c=1.525$ for $\Phi=0.45$)
 as the $1/\be_f\tend{0}$ limiting case of the frozen DHS model. On this perfect lattice,
from the behavior of the Binder cumulant, we obtain a PM/FM transition at $T^*_c=0.81$ with a long range (LRO) 
FM order at low temperature. This latter point is deduced from the independence with $N$ at $T^{*}<0.45$ 
of the moments $m_k$, $k=1,2,4$. Notice that in opposite to what is found on the dipolar Ising 
model~\cite{fernandez_2000} (see below) $T^*_c$ is much higher than on the BCC ($a/c=1$) lattice for the same 
value of $\Phi$ ($T^*_c(BCC,\phi=0.45)=0.345\pm0.015$~\cite{russier_2022}) which clearly results from the two 
independent degree of freedom per moment in the freely rotating dipoles conversely to the DIM case.
\\
Then we consider finite values of $\be_f$ still larger than the PM/FM transition inverse temperature of the DHS,
between 3.80 and 5.71, 
(see Table~\ref{tab_fr}) chosen as they sample the lower half of the inverse temperature range where 
$\la_s$ linearly increases from its vanishing limit~(figure~(\ref{dhs_45_lam_st})). 
From the result displayed in figure~(\ref{dhs_45_lam_st}) we deduce that the anisotropy of the DHS structure 
vanishes below $\be_f\simeq{}3.70$, and accordingly we conclude that 
for $\be_f<3.70$, the features of the frozen DHS will coincide with those of $\be_f\ll{}1/T^*_c(DHS)$ limiting 
case and a PM/SG transition at $T^*_c=0.23\pm{}0.015$ is expected. \\
At both $\be_f=3.8$, and 4.0, 
following the same protocol as the one used above for the frozen DHS with the isotropic
HS structure, we get a PM/SG transition, at 
$T^*_c=0.29\pm{0.02}$ and $0.345\pm{0.02}$ respectively. 
Then we have performed the calculations of the Binder cumulants $B_m(T^*,N)$ and $B_{md}(T^*,N)$ on the one hand and
of the heat capacity $C_v$ and the magnetic susceptibility $\chi_M$ on the other hand for $\be_f\geq{}4.25$. 
In figure~(\ref{dipfr_045_be45_m_bm}) the magnetization $m_1$ and the Binder cumulant are displayed in the case $\be_f=4.5$; the 
system presents qualitatively the same behavior for the other values of $\be_f\ge{}4.25$. 
A PM/FM transition is found for all values of $\be_f\geq{}4.25$. 
The heat capacities are compared for $\be_f=5.71$ and 3.80 on figure~(\ref{dipfr_045_cv_be571-380}) 
as examples of the PM/FM and PM/SG transitions respectively. Both the finite size effect and the lambda-like shape 
of $C_v$ are observed as expected only in the former case.
In this system, as is the case for the dipoles on FCC lattice, 
we expect a slightly negative value for the $C_v$ exponent $\al$ and therefore a non singular $C_v$. 
This is deduced from the relation $\al=2-d\nu$ and the scaling behavior we get for the Binder cumulant $B_{md}$
which seems in agreement with the 3D Heisenberg case ($\nu=0.707$) for the DHS and the dipoles on FCC lattice 
($\nu=0.692$)~\cite{bruce_1974} for the frozen DHS at the four values of $\be_f\geq{}4.25$ considered. 
Nevertheless as already mentioned, the determination of the critical exponents is beyond the scope of the 
present work.
From figures~(\ref{lnm2_lnl},\ref{1_moins_bm}), we clearly see the different behaviors of both $m_2$ and 
$(1-B_{md})$ with $N$ when going from the PM/SG ($\be_f=3.8$) to the PM/FM ($\be_f=5.71$) transitions
regions of the phase diagram. Specifically, $m_2$ decreases and $(1-B_{md})$ increases with $N$ whatever 
the value of $T^*$ for $\be_f=3.8$ while $(1-B_{md})$ decreases with $N$ below $T^*_c$ for $\be_f=5.71$.
At $\be_f=5.71$ only a very small variation of $ln(m_2)$ as a function of $ln(N^{1/3})$ persists below $T^*_c$ 
(see figure~(\ref{lnm2_lnl}b)). In any case, beyond $T^*\sim{}0.8$ we recover the perfect PM dependence, 
namely $m_2\sim{}1/L^3$ whatever the value of $\be_f$. 
\\ \indent
When approaching the SG/FM line, at $\be_f=4.25$ (not shown), the moments $m_k$ at low temperature decrease with $N$ at least for $N\le{}1177$, 
and we may expect a FM quasi long range order (QLRO) as already evidenced on related systems
with isotropic structure~\cite{russier_2020b,alonso_2020b}. 
The actual signature of the FM QLRO namely 
a diverging behavior of the magnetic susceptibility $\chi_m$ at low temperature 
(see the remark below in section~\ref{dipfr_26} and figure~(\ref{fr_dhs_026_m_xm_bm}b) for the typical 
low temperature behavior of $\chi_m$)
with increasing $N$, and a finite value of $(1-B_{md})$ at the thermodynamic limit,
have not been obtained.
As a result, we conclude that no FM-QLRO takes place in the FM region of the ($T^*,\be_f$) phase diagram.
\\ \indent
Finally we have located the SG/FM line below the PM/FM and the PM/SG lines from the 
finite size behavior of the $B_{md}$ curves in terms of $\be_f$ at constant temperature as follows. 
In the FM region, $B_{md}$ is increasing with the system size, while this is the opposite in the SG region. 
Hence the SG/FM line is determined from the crossing point of the $B_{md}$ curves in terms of $\be_f$ 
at constant temperature. To this aim, we have taken the values $\be_f=3.80$ and 4.25 as bracketing values,
since we found that the PM/FM line transforms in a PM/SG line in the very vicinity of $\be_f=4.0$.
We find the SG/FM line located at $\be_{fc}=4.19\pm{}0.03$ for $T^*$ in the range $0.11\leq{T^*\leq}0.20$
and at $\be_{fc}=4.17\pm{}0.04$ for $T^*$ in the range $0.22\leq{T^*\leq}0.26$.
Given the uncertainty bar, we cannot conclude on a $T^*$ dependence of $\be_{fc}$ and accordingly on a re-entrance 
behavior.
\\ \indent
The corresponding phase diagram in the ($T^*,\be_f$) plane is displayed in figure~(\ref{diag_ph_dipfr_045}~a).
In this phase diagram, we did not locate precisely the tricritical point where the PM, FM and SG phases
meet together; however, its $\be_f$ value 
can be bracketed first in between the largest (smallest) value of the $\be_f$ values considered on the 
SG/PM (FM/PM) lines and second from the continuation of the SG/FM line with the result 
$\be_M\in[4.10,4.22]$.
\\ \indent
\begin{table}[h!]
\tab { | c | c |c | c | c | }
\hline
 $\be_f$ & $P_2$ & $\s$ & $T_c^{*(a)}$ & $T_c^{*(b)}$ \\
\hline
 $\infty$   ~&~ 	 ~&~ 	   ~&~ 0.810 $\pm$ 0.015 ~&~ 2.70 $\pm$ 0.02  \\
 5.71  ~&~ 0.404 ~&~ 0.566 ~&~ 0.461 $\pm$ 0.02  ~&~ 2.51 $\pm$ 0.03 \\
 5.00  ~&~ 0.302 ~&~ 0.657 ~&~ 0.429 $\pm$ 0.02  ~&~ 2.42 $\pm$ 0.03 \\
 4.50  ~&~ 0.212 ~&~ 0.757 ~&~ 0.395 $\pm$ 0.02  ~&~ 2.34 $\pm$ 0.02 \\
 4.25  ~&~ 0.163 ~&~ 0.825 ~&~ 0.371 $\pm$ 0.02  ~&~ 2.25 $\pm$ 0.03 \\
 4.00  ~&~ 0.101 ~&~ 0.941 ~&~ {\it 0.345 $\pm$ 0.02}  ~&~ 2.17 $\pm$ 0.03 \\
 3.80  ~&~ 0.059 ~&~ 1.063 ~&~ {\it 0.29 $\pm$ 0.020}  ~&~ {\it 2.11$\pm$ 0.020} \\
 0     ~&~ ~&~ ~&~ {\it 0.23 $\pm$ 0.015} ~&~ {\it 0.60 $\pm$ 0.1} \\
 \hline
 \bat
 \caption {\label{tab_fr}
 Values of $\be_f$ considered for the frozen DHS and frozen DIM models at $\Phi=0.45$.
 $P_2$ is the value of the nematic order parameter of the DHS at $T^*=1/\be_f$ and $\s$ is the 
corresponding variance of the easy axes distribution for the DIM model (see text in 
section~\ref{dim_045}).
 $T_c^{(a),(b)}$ are the values of $T^*_c$ for the frozen DHS and DIM models respectively.
 $\be_f=\infty$ refers to the frozen BCT structure (see text).
 $\be_f=0$ refers to the isotropic HS like structure corresponding to the frozen
 structure of the DHS in the high temperature limit.
 We indicate by italics the PM/SG transitions.
  }
\end {table}
\FloatBarrier

 \subsubsection {Frozen dipolar Ising model ($\la_u\tend\infty$). $\Phi=0.45$}
 \label {dim_045}
 The frozen DIM at $\Phi=0.45$ has been considered for the same values of $\be_f$ as the frozen DHS, 
 listed in Table~\ref{tab_fr}.
Here, in addition to the structure anisotropy which is still quantified by the values of the 
structural nematic order parameter $\la_s(\be_f)$ deduced from the DHS, we have to characterize the distribution of 
easy axes. This is done by the nematic order parameter $P_2$ of the DHS since the $\hat{n}_i$ are taken equal 
to the $\hat{\mu}_i$ of the DHS at $T^*=1/\be_f$. 
The values of $P_2(\be_f)$ are given in Table~\ref{tab_fr}.
These non vanishing values of $P_2$ can be translated as the texturation of the easy axes 
distribution, by representing the latter by a probability distribution $P(\theta)$ 
of the polar angles $\theta_i=Acos(\hat{n}_i\ldotp\hat{d})$. Using 
$P(\theta)=C.sin(\theta)(exp(-(\theta)^2/2\s^2)+exp(-(\pi-\theta)^2/2\s^2))$~\cite{alonso_2019}, 
the values that we deduce for the variance $\s$ of the easy axes distribution range between 
$\s=0.566$ and 0.941 (see Table~\ref{tab_fr}) 
while the random distribution is obtained for $\s\ge\pi/2$. 
Therefore in the frozen DIM, the disorder brought by the MAE is limited by the texturation 
of the axes distribution, which increases when $\be_f$ increases. In Ref.~\cite{alonso_2019} we found that 
the DIM with an isotropic frozen distribution at $\Phi=0.64$ orders at low temperature in a FM phase for 
$\s\leq{}0.53$ and a SG phase otherwise. Hence, we can expect that the disorder introduced by the MAE 
in the frozen DIM considered here at $\Phi=0.45$ is not sufficient to make the FM transform in a SG phase 
at least for $\be_f=5.71$.
 
\begin {figure} [h] 
   \includegraphics [width = 0.90\textwidth , angle = -00.00]{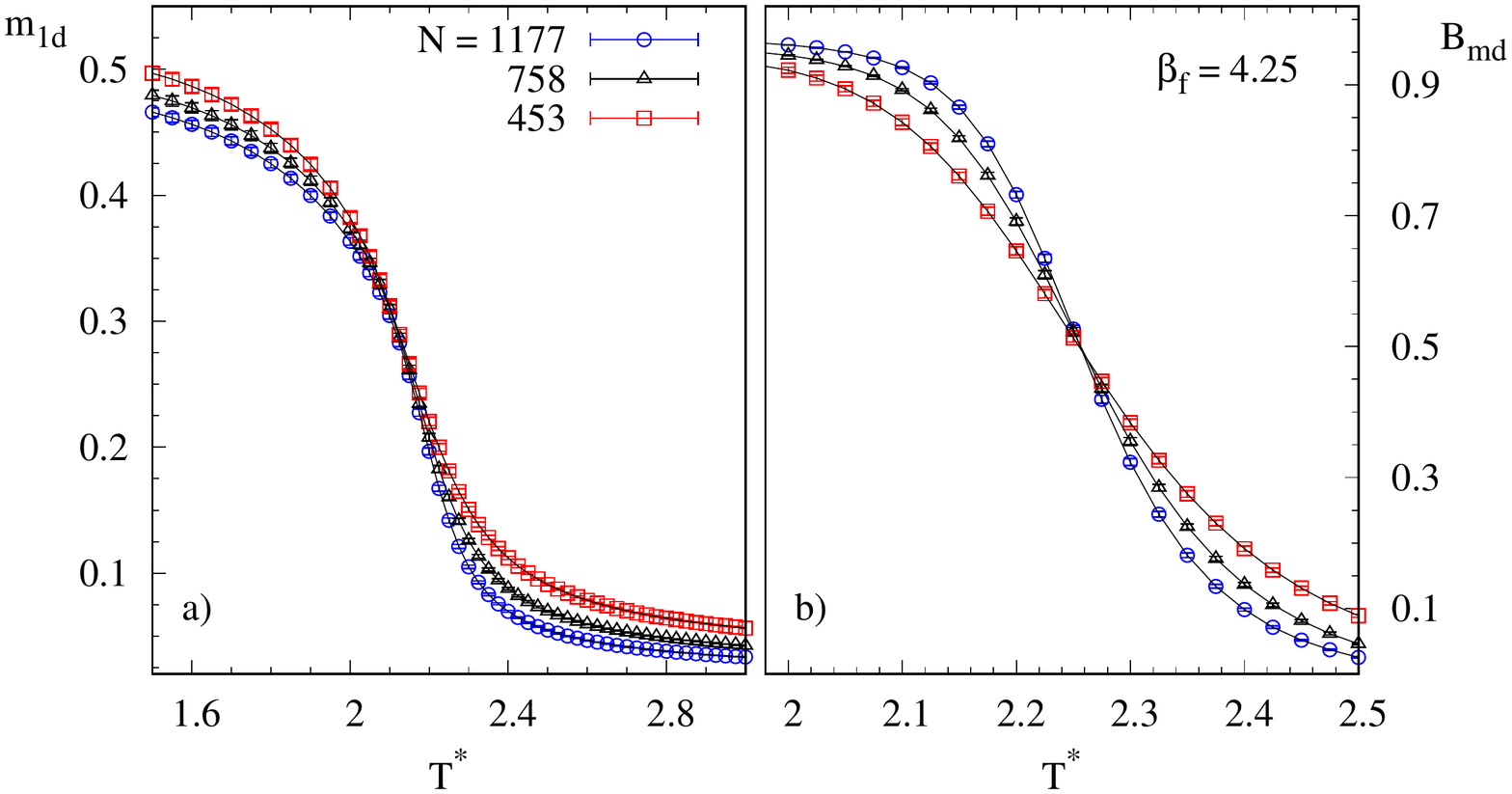}
   \vskip -0.03\textheight
   \caption {\label {m_bm_dim_045}
   Frozen DIM model at $\Phi=0.45$ and $\be_f=4.25$. 
   a) Magnetization $m_{1d}$ and b) Binder cumulant $B_{md}$ in terms of $T^*$.}
   \end {figure}

As expected, the frozen DIM at $\be_f=5.71$ presents a PM/FM transition; this is also the case at the 
lower freezing inverse temperatures studied including $\be_f=4.0$. 
We show on figure~(\ref{m_bm_dim_045}), as a typical example representative of the whole set of freezing 
temperatures studied here the magnetization and the Binder cumulant $B_{md}$ for $\be_f=4.25$
for the system sizes $N=453$, 758 and 1177.  
The features of the five cases of the frozen DIM model are 
qualitatively similar, namely a lambda-shape of the $C_v$ curves with a marked system size behavior,
a pronounced peak and also a marked system size dependence on the $\chi_m$ curves, and finally a 
crossing point in the $B_{md}$ (see figure~(\ref{m_bm_dim_045})) curves corresponding to different 
values of $N$, from which the transition temperature $T^*_c$ is deduced (see Table~\ref{tab_fr}).
At $\be_f=3.8$ the frozen DIM orders in a SG phase whose transition temperature is determined from the crossing
point of the reduced spin-glass correlation length $\xi/L$ curves corresponding to the 3 system sizes.
We have also determined the FM/SG line below the PM/FM and the PM/SG lines. As is the case for the
frozen DHS we do not find evidence of a reentrance behavior and the SG/FM line is located at
$\be_{fc}=3.97\pm{}0.02$ with however a greater uncertainty ($\pm{}0.07$) in the very vicinity of the SG/PM transition.
When $\be_f<3.70$ where the structural anisotropy vanishes, the frozen DIM is expected to order in
a spin-glass phase, and the corresponding $T^*_c$ then coincides with the one obtained with the 
isotropic hard sphere like distribution~\cite{alonso_2020b} from the spin-glass Binder cumulant $B_{sg}$
and the reduced spin-glass correlation length $\xi/L$, given in Table~\ref{tab_fr}.
In the phase diagram of the frozen DIM in the ($T^*,\be_f$) plane, shown in 
figure~(\ref{diag_ph_dipfr_045}~b) the PM/FM line is located at higher reduced critical temperatures,
and depends less on the value of $\be_f$. This comes from the additional source of anisotropy brought by 
the non isotropic distribution of Ising axes.

 \subsubsection {Frozen dipolar hard sphere and dipolar Ising models at $\Phi=0.262$}
 \label {dipfr_26}
\begin {figure} [h] 
   \includegraphics [width = 0.90\textwidth , angle =  00.00]{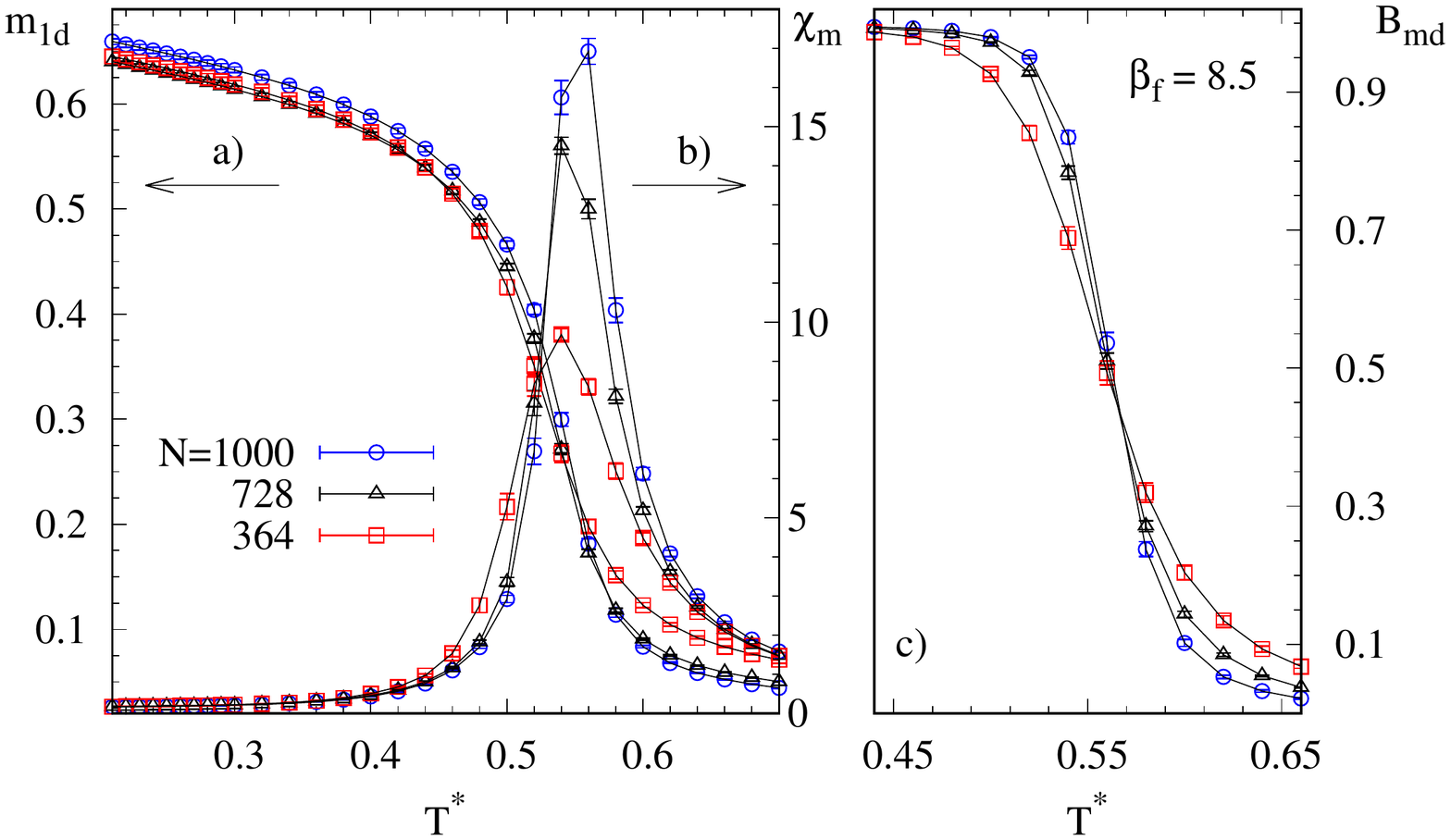}
   \vskip -0.03\textwidth
   \caption {\label {fr_dhs_026_m_xm_bm}
   Frozen DHS model at $\Phi=0.262$ and $\be_f=8.5$.
   a) Magnetization $m_{1d}$, b) magnetic susceptibility $\chi_m$ and c) Binder cumulant $B_{md}$. 
   }
   \end {figure}

In the low density case, $\Phi=0.262$, we consider the frozen DHS and the frozen DIM at the inverse frozen 
temperature $\be_f=8.5$.
 According to the note~\cite{note_tphi} the location of this point in the frozen DHS $(\be_f,T^*)$ phase diagram 
 relative to the DHS PM/FM transition and the onset of the structural anisotropy (see section~\ref{liq_dhs_045})
 may be compared qualitatively to that of the frozen DHS at $\Phi=0.45$ and $\be_f\sim{}5.0$. 
The system sizes used are $N=364$, 728 and 1000.
The results for the magnetization, the magnetic susceptibility $\chi_m$ and the magnetic Binder cumulant $B_{md}$ 
are shown in figure~(\ref{fr_dhs_026_m_xm_bm}).
First of all, from the finite size behavior of the Binder cumulant $B_{md}$ (see figure~(\ref{fr_dhs_026_m_xm_bm}c)), 
we conclude that the system presents a PM/FM transition at $T^*_c=0.56\pm0.02$. 
The $C_v$ and $\chi_M$ curves with a strong finite size dependence and a pronounced peak in the vicinity of 
$T^*_c$ (see figure~(\ref{fr_dhs_026_m_xm_bm}b)) corroborate the PM/FM nature of the transition. 
It is worth noticing that the low temperature behavior of $\chi_M$ 
(figure~(\ref{fr_dhs_026_m_xm_bm}b)) 
is qualitatively similar to that obtained on both the frozen DHS at $\Phi=0.45$ and $\be_f\geq{}4.25$ and 
the frozen DIM at $\Phi=0.45$ and $\be_f\geq{}4.0$.
Notice that the limiting point ($\be_f\tend{0}$) of the frozen DHS at $\Phi=0.262$  corresponds to the PM/SG transition of the 
frozen system with isotropic hard sphere like structure with $T^*_c\simeq{}0.12$~\cite{alonso_2020b}.
From the comparison of $T^*_c$ between the frozen DHS at $(\Phi=0.262,\be_f=8.5)$ and $(\Phi=0.45,\be_f\sim{}5)$ we see that 
an important difference is the much larger deviation of $T^*_c$ 
with respect to the PM/SG line expected for $\be_f$ beyond the onset of structural anisotropy.
This is likely due to the more efficient dipolar interaction along the nematic direction compared to its 
transverse component when the volume fraction decreases, for a given structural anisotropy. 
\\

\begin {figure} [h] 
   \includegraphics [width = 0.90\textwidth , angle = -00.00]{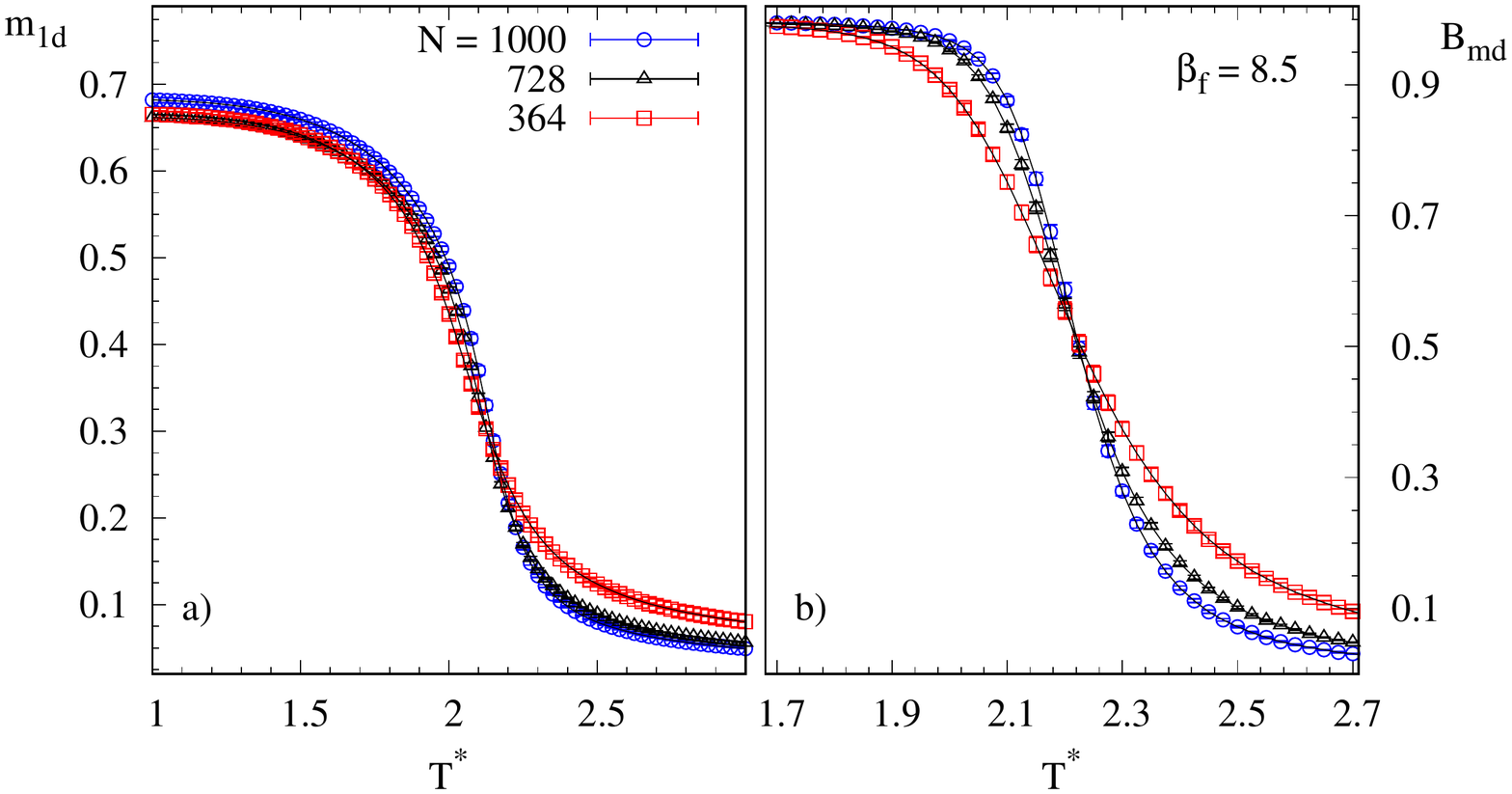}
   \vskip -0.03\textheight
   \caption {\label {m_bm_dim_026}
   Frozen DIM model at $\Phi=0.262$ and $\be_f=8.5$. 
   a) Magnetization $m_{1d}$ and b): Binder cumulant $B_{md}$ in terms of $T^*$.}
   \end {figure}

%
 The frozen DIM model at $\Phi=0.262$ and $\be_f=8.5$ whose magnetization $m_{1d}$ and Binder cumulant $B_{md}$ are
shown on figure~(\ref{m_bm_dim_026}), is shown to present a clear FM/PM transition at a $T^*_c=2.25\pm0.05$. 
This is deduced as above from the crossing point behavior of the magnetization Binder cumulant 
(see figure~(\ref{m_bm_dim_026})) the strong finite size dependence of both the heat capacity, $C_v$ and the magnetic
susceptibility, $\chi_M$. 
Conversely to the case of frozen DHS model for a given structural anisotropy, here the value of $T^*_c$ is lower than although  
very close to the one obtained at $\Phi=0.45$. 
This weaker influence of the volume fraction is a consequence of the fact that the moments are imposed along the Ising axes
whose distribution is fixed through the structural anisotropy which suppresses one degree of freedom per moment. 

 \section {Conclusion}
 \label   {conclusion}

 In this work we have determined the phase diagram of an ensemble of dipolar hard spheres
with or without a uniaxial anisotropy. In the latter case the infinitely strong anisotropy
was considered, where the system transforms in a dipolar Ising model.
Besides its fundamental interest such a model is useful to understand the dipolar effects
in any case present in magnetic nanoparticles assemblies in the single domain regime, occurring for
nanoparticles under a critical size.
The principal motivation of the present work is to focus on systems with frozen structures
presenting an anisotropy, which means that along a preferential direction the nearest neighbor
distance is smaller than its average value. Moreover, we get this structural anisotropy 
starting from the liquid DHS in its polarized state ($i.e.$ for inverse temperatures 
$\be_f>\be_c(DHS)$ where $\be_c(DHS)$ is the critical inverse temperature of the DHS at 
the volume fraction considered). As a result
the anisotropy, quite naturally quantified from the nematic structural order parameter,
can be tuned at will, and either $\la_s$ or $\be_f$ is a control parameter which
quantifies the deviation from the totally disordered structure. We are then faced with
the determination of the phase diagram in the $(\be_f,T)$ plane or equivalently in terms of 
the amount of disorder. 
\\ \indent
The first and important result of this work is that the structural 
anisotropy even at small values of $\la_s$, or equivalently for $\be_f$ close to $\be_c(DHS)$
the frozen DHS orders at low temperature in a FM phase. We emphasize that this result
holds for volume fractions smaller than the threshold value under which the same system but 
with disordered and isotropic frozen structure presents only a spin-glass phase at low temperature.
The low temperature FM phase is thus obtained down to the low volume fraction, $\Phi=0.262$.
It is worth mentioning that another way to reduce he dipolar volume fraction 
instead of a simple dilution of the pure DHS fluid
is to substitute part of the dipolar hard spheres by non magnetic ones.
Doing this increases the ferromagnetic transition
temperature ($T^*_c(DHS)$) of the DHS fluid~\cite{malherbe_2021}. 
As a result, higher freezing temperatures of dilute DHS should be considered.
\\ \indent
The second important result is that the frozen DIM with the same structure orders more easily 
in a FM phases than the frozen DHS. We emphasize that in the DIM, the Ising axes also
are textured in the frozen structure as they follow the moments distribution of the 
initial liquid DHS. \\ \indent
Finally we note that the usefulness of the DHS systems phase diagrams in the field of magnetic
nanoparticles (MNP) research concern mainly the study of the DDI effect on the magnetic properties
of MNP assembled in superstructures and/or concentrated ferro-fluids. In this framework,
this work may suggest a way to get the so-called super-FM phase induced by DDI
from the synthesis of structurally textured MNP organization.

 \section {Acknowledgements}
 This work was granted access to the HCP resources of CINES under allocations
 2021-A0100906180 and 2022-A0120906180 made by GENCI, CINES, France.
 J.J.A. thanks SCBI at University of M\'alaga for additional computer time.
%

  \newpage
%
 %
 %
 %

\begin{thebibliography}{10}
\bibitem{note_bct}
For a BCT structure with axis $c=\hat{z}$, $c=1$ and lattice constant $a$ in
  the plane normal to the $c$ axis, the projections of the distances between
  neighbors in the planes normal to the axis $c$ are ($a, a\sqrt{2}, 2a,
  a\sqrt{5}, a\sqrt{8},...$) and ($a/\sqrt{2}, a\sqrt{10}/2, 3a/\sqrt{2},
  \sqrt{26}a/2, a\sqrt{17/2},...$) in the central and in the median planes,
  $i.e.$ located at $z=0$ and $z=c/2$ respectively.
\bibitem{note_tphi}
As a consequence of the $1/r^3$ dependence of the DDI, one expects the behavior
  of the system at different values of the volume fraction $\Phi$ to depend on
  the temperature through $T^*/\Phi$ instead of $T^*$. This holds for instance
  for the value of the PM/FM transition temperature of the DHS. However, it
  must be emphasized that such a mapping relates upon an averaged effect of the
  DDI on the ensemble of particles assumed to be smeared out in the volume and
  does not hold once the structure becomes significantly anisotropic.
\end{thebibliography}
\end   {document}